\documentclass[12pt]{iopart}

\usepackage{subfigure}

\usepackage{graphicx}
\usepackage{mathrsfs}
\usepackage[british]{babel}
\usepackage{enumerate}
\usepackage{enumitem}
\usepackage{bbm}

\usepackage[sort&compress,numbers]{natbib}
\usepackage{doi}
\usepackage{hyperref}

\usepackage{comment} 
\expandafter\let\csname equation*\endcsname\relax 
\expandafter\let\csname endequation*\endcsname\relax 

\usepackage{amsbsy,amssymb,amsmath,bm}
\usepackage{graphicx,color,epsfig,rotate}

\catcode`@=11 
\renewcommand\tableofcontents{%
  \section*{\contentsname}%
  \@starttoc{toc}%
}
\catcode`@=12

\def\XXint#1#2#3{{\setbox0=\hbox{$#1{#2#3}{\int}$}
     \vcenter{\hbox{$#2#3$}}\kern-.5\wd0}}

\newcommand{\ket}[1]{\left|#1  \right>}

\renewcommand{\imath}{i}

\newcommand{\beq}{\begin{equation}}
\newcommand{\eeq}{\end{equation}}

\usepackage{xcolor}

\usepackage{multirow}
\newcommand{\be}{\begin{equation}}
\newcommand{\ee}{\end{equation}}

\usepackage{etoolbox}

\makeatletter
\def\@mkboth#1#2{}
\newlength\appendixwidth
\preto\appendix{\addtocontents{toc}{\protect\patchl@section}}
\newcommand{\patchl@section}{%
  \settowidth{\appendixwidth}{\textbf{Appendix }}%
  \addtolength{\appendixwidth}{1.5em}%
  \patchcmd{\l@section}{1.5em}{\appendixwidth}{}{\ddt}%
}
\makeatother

\newcommand\particle{\uparrow}
\newcommand\hole{\downarrow}

\begin{document}


\title[Anomalous transport from quasiparticles]{Anomalous transport from hot quasiparticles in interacting spin chains}

\author[]{Sarang Gopalakrishnan}
\address{Department of Physics, The Pennsylvania State University, University Park, PA 16802, USA}

\address{Department of Electrical and Computer Engineering, Princeton University, Princeton, NJ 08544, USA}

\author[]{Romain Vasseur}
\address{Department of Physics, University of Massachusetts, Amherst, MA 01003, USA}

\ead{rvasseur@umass.edu}
\vspace{10pt}

\begin{abstract}

Many experimentally relevant quantum spin chains are approximately integrable, and support long-lived quasiparticle excitations. A canonical example of integrable model of quantum magnetism is the XXZ spin chain, for which energy spreads ballistically, but, surprisingly, high-temperature spin transport can be diffusive or superdiffusive.  We review the transport properties of this model using an intuitive quasiparticle picture that relies on the recently introduced framework of generalized hydrodynamics. We discuss how anomalous linear response properties emerge from hierarchies of quasiparticles both in integrable and near-integrable limits, with an emphasis on the role of hydrodynamic fluctuations. We also comment on recent developments including non-linear response, full-counting statistics and far-from-equilibrium transport. We provide an overview of recent numerical and experimental results on transport in XXZ spin chains.

\end{abstract}

\setcounter{tocdepth}{2}
\tableofcontents

\section{Introduction}

A natural way to study a complex system, such as the fluid of electrons in a metal, is to perturb it and see how it responds.
Of the tools an experimentalist has to perturb the electron fluid, the simplest is to apply an electric field; the corresponding response is the (electrical) conductivity. Since the conductivity is so simple to measure, it is the most-studied dynamical signature of condensed-matter systems. Historically, these studies focused on the low-temperature regime, for at least three reasons. First, the conductivity in this limit can be treated in terms of a dilute gas of elementary excitations, allowing for controlled theoretical calculations. Second, heating a metal to its Fermi temperature causes it to interact strongly with lattice vibrations and other extraneous degrees of freedom; at low temperatures these freeze out, and we can probe the intrinsic behavior of the electron liquid. Third, and most importantly, low temperatures are where one sees qualitatively striking dynamical phenomena like superconductivity---by contrast, high-temperature states are expected to be normal metals with short mean free paths. Equivalently (by the Einstein relation) we expect high-temperature transport to be diffusive.

\begin{figure}[tb]
\begin{center}
\includegraphics[width = 0.85\textwidth]{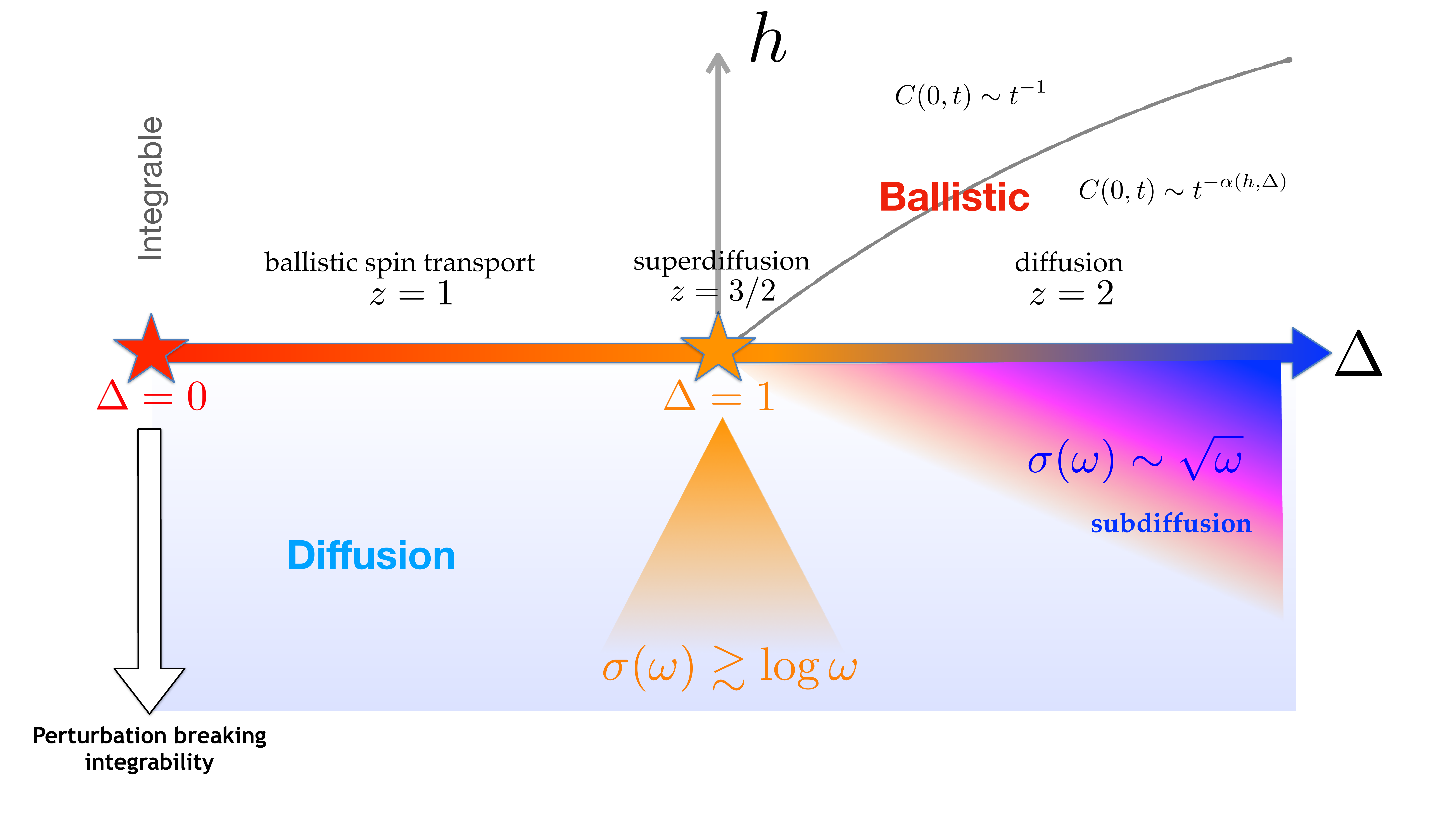}
\caption{{\bf Summary phase diagram of spin transport in the XXZ spin chain}. In the integrable limit and at half-filling (zero net magnetization, magnetic field $h=0$), spin transport is ballistic in the easy-plane ($\Delta<1$) phase, diffusive in the easy-plane phase ($\Delta>1$) with anomalous full counting statistics, and superdiffusive with $z=3/2$ at the isotropic point $\Delta=1$. Away from half-filling $h>0$, spin transport is ballistic for all $\Delta$ but the local relaxation of the dynamical structure factor is anomalous with continuously exponents. Away from integrability, for weak perturbations and at half-filling, transport is likely diffusive everywhere at long times. However, at large anistropy, spin transport is subdiffusive with $z=4$ for a parametrically large regime. For $\Delta=1$, perturbations preserving the $SU(2)$ symmetry perturbatively lead to superdiffusive transport due to long-lived quasiparticles.   }
\label{FigSummary}
\end{center}
\end{figure}

Hydrodynamics supports this empirical conclusion. Imagine initializing a system in a non-equilibrium initial state. By locality, at short times one can divide the system up into local mesoscale ``cells.'' Generic many-body dynamics is chaotic, so after a short evolution time the local state in each cell approaches a random state subject to the relevant conservation laws (which apply locally). Since all other information about the initial state has been lost, the subsequent evolution just involves the equilibration of these few remaining slow variables across the system. In the limit of long-wavelength fluctuations, one can write down the most general equations for these relaxation processes within a gradient expansion. 
Whether this remaining dynamics is rich or simple depends on the nature of these residual conserved quantities. When momentum is conserved, the resulting fluid dynamics is rich; when it is not---as in electronic systems embedded in a crystal lattice or an amorphous medium---hydrodynamics predicts that all transport should be governed by the diffusion equation. At low temperatures, in two or more dimensions, a system might acquire additional slow modes, such as long-wavelength order-parameter fluctuations in a phase with spontaneous symmetry breaking~\cite{kirkpatrick2002long}. But at high temperatures, diffusion is the inevitable consequence of standard hydrodynamics, and in one dimension all nonzero temperatures are high enough to prevent ordering. 

A series of experimental advances around the year 2000, most notably in ultracold atomic gases, allowed one to experimentally access the regime of hot but isolated quantum matter~\cite{pssv}. These studies led to a realization that is obvious in retrospect but has consequences that had not been fully appreciated:  the assumption of rapid chaotic local thermalization is false in many one-dimensional systems, invalidating the logic we sketched in the previous paragraph. Rather, on experimentally relevant timescales, one-dimensional and quasi-one-dimensional quantum gases are approximately integrable. Integrable systems can be regarded as having extensively many conserved charges, or equivalently as hosting stable, ballistically propagating quasiparticles. This structure suggests that the appropriate framework for describing them is not hydrodynamics but a collisionless Boltzmann equation. Such a description was recently developed using the framework of generalized hydrodynamics (GHD)~\cite{Doyon, Fagotti}. 
GHD offers a quantitative description of transport from nonequilibrium initial states~\cite{Doyon, Fagotti, SciPostPhys.2.2.014,piroli2017, 2017arXiv171100873C, DOYON2018570, doyon2017dynamics, PhysRevLett.120.176801,PhysRevB.97.081111,BBH0,PhysRevLett.119.195301,bertini2018low,PhysRevB.98.075421,solitongases,2019arXiv190601654B, 10.21468/SciPostPhys.8.3.041,PhysRevB.100.035108,PhysRevB.96.220302,10.21468/SciPostPhys.8.1.007,DoyonMyers,DoyonToda,Bulchandani_2019Toda, Cao_2019Toda,ruggiero2019quantum,Balasz,PhysRevLett.125.070602,10.21468/SciPostPhys.8.2.016,PhysRevE.102.042128,PhysRevB.101.035121,PhysRevB.103.035130,PhysRevLett.125.240604}, entanglement dynamics~\cite{alba2017entanglement,Bertini_2018,10.21468/SciPostPhys.7.1.005} and correlation spreading~\cite{10.21468/SciPostPhys.5.5.054,10.21468/SciPostPhysCore.3.2.016}, as well as analytical expressions for linear response quantities such as Drude weights~\cite{PhysRevLett.119.020602,BBH, GHDII, PhysRevB.96.081118, PhysRevB.97.081111} and diffusion constants~\cite{dbd1, ghkv, dbd2,GV19,10.21468/SciPostPhys.9.5.075,doyon2019diffusion}. As we will discuss in detail in this review, the GHD framework also revealed the existence of anomalous transport regimes in strongly interacting spin chains~\cite{PhysRevLett.106.220601,lzp,idmp,GV19,PhysRevLett.123.186601,gvw, dupont_moore,vir2019, 2019arXiv190905263A,PhysRevLett.122.210602,PhysRevB.102.115121,2020arXiv200908425I,dmki,PhysRevLett.125.070601}, which  motivated recent experiments~\cite{Jepsen:2020aa,bloch2014,2020arXiv200913535S,wei2022quantum}. 


Concurrently with these advances, it also became possible to perform efficient numerical simulations of the dynamics of lattice models using matrix-product methods~\cite{schollwoeck}. These simulations at first focused on the XXZ spin chain, governed by the Hamiltonian 
\begin{equation}
\hat H_{\mathrm{XXZ}} = J\sum\nolimits_i \left( \hat S_i^x \hat S_{i+1}^x + \hat S_i^y \hat S_{i+1}^y + \Delta \hat S_i^z \hat S_{i+1}^z \right),
\end{equation}
where $\hat S^\alpha=\hat \sigma^\alpha/2$ are spin-$1/2$ operators, and we shall set $J=1$ in what follows. The total magnetization along the $z$ axis, $\sum\nolimits_i \hat S^z_i$, is conserved under $\hat H_{\mathrm{XXZ}}$. We will be interested in how magnetization is transported. Note that $\hat H_{\mathrm{XXZ}}$ maps onto a model of interacting fermions via a Jordan-Wigner transformation~\cite{sachdev2011}. In the fermion language, the total magnetization maps onto the total particle number.

The XXZ model has a ground-state phase transition at $|\Delta| = 1$. For $|\Delta| < 1$, spins in the ground state point along the equator of the Bloch sphere, while for $|\Delta| > 1$ they point toward the poles. The nature of the elementary excitations is also distinct: in the former case they are long-wavelength, linearly dispersing spin waves and the spectrum is gapless, and in the latter they are domain walls and the spectrum is gapped.
One might expect these ground-state distinctions to be irrelevant for high-temperature transport: 
regardless of $\Delta$, 
the model is integrable and has ballistic quasiparticles. 
Naively this suggests that transport should always be ballistic. 
Well before the advent of GHD, however, it had been appreciated that ballistic quasiparticles can give rise to diffusive transport of some charges~\cite{PhysRevLett.78.943, PhysRevLett.95.187201, PhysRevB.57.8307}.
What numerics revealed is that \emph{both} forms of transport occur in the XXZ model, depending on the value of $\Delta$:
at $|\Delta| = 1$, transport undergoes an \emph{infinite-temperature phase transition} from ballistic to diffusive (Fig.~\ref{FigSummary}). The critical point $\Delta = 1$ exhibits superdiffusion~\cite{PhysRevLett.106.220601,lzp} with an anomalous transport exponent $x \sim t^{2/3}$ (which is analytically understood); based on numerical evidence from scaling functions~\cite{PhysRevLett.122.210602}, it is also believed that transport is governed by the Kardar-Parisi-Zhang equation~\cite{kpz} (though this has not yet been microscopically derived). The recognition that such high-temperature dynamical phase transitions can occur even in one dimension is one of the central new discoveries in quantum dynamics. (A slightly earlier instance is the many-body localization transition in disordered spin chains~\cite{RevModPhys.91.021001}.)

The aim of this review is to describe our current understanding of why distinct dynamical phases can occur in integrable spin chains. The explanation involves two crucial ingredients: the existence of \emph{stable quasiparticles} at \emph{high density}. 
Because the quasiparticles are stable at all temperatures, a change in their character has consequences for dynamics at all temperatures. (A related manifestation of this fact is the existence of ``strong zero modes''~\cite{fendley2016strong}.)
At high densities, collisions among quasiparticles renormalize the properties of each individual quasiparticle; these renormalization effects are key to the possibility of non-ballistic transport in integrable systems. 
When the nature of the quasiparticles changes, it is natural for these renormalization effects also to change.

\subsection{Scope and organization of this review}

This review aims to be a self-contained treatment of anomalous dynamics in integrable and near-integrable systems. We do not assume any knowledge of background topics such as generalized hydrodynamics or integrability-breaking; nor do we offer a comprehensive introduction to these topics. Instead, we introduce and physically motivate key results from these areas when they are relevant to our discussion. For recent pedagogical introductions to generalized hydrodynamics see~\cite{doyon2019lecture,Bastianello_2022,Bouchoule_2022}; for diffusive corrections to GHD, see~\cite{De_Nardis_2022}; for an introduction to integrability-breaking perturbations within GHD see~\cite{Bastianello_2021}. The review article~\cite{bertini2020finite} has an extensive discussion of numerical and exact results on transport. 

Our work has some overlap with a recent review on superdiffusion in integrable systems~\cite{Bulchandani_2021}. Our emphasis is different, however: Ref.~\cite{Bulchandani_2021} focused on superdiffusion at the Heisenberg point; our work focuses more generally on the nontrivial consequences of long-lived quasiparticles for transport and dynamics. This new emphasis is motivated by recent developments (mostly since the publication of Ref.~\cite{Bulchandani_2021}) showing that superdiffusion is only one of a range of surprising transport phenomena that stem from the existence of long-lived quasiparticles. In addition, transport in XXZ spin chains has become a topic of great experimental relevance since the publication of Refs.~\cite{bertini2020finite, Bulchandani_2021}. Accordingly, this work offers a more detailed survey of the current experimental status of the field.

The rest of this review is organized as follows. In Sec.~\ref{SecBackground} we give a brief self-contained introduction to the concepts that are needed to understand GHD and transport in integrable (and nearly integrable) systems. In Sec.~\ref{SecFoldedXXZ} we provide a detailed discussion---both in the linear-response regime and away from it---of the dynamics of one of the simplest examples with anomalous behavior, the ``folded'' limit of the XXZ spin chain. In Sec.~\ref{SecEasyAxis} we extend our results away from the folded limit to the easy-axis regime of the XXZ spin chain. In Sec.~\ref{SecXXX} we extend them further to the isotropic Heisenberg point, at which we find very different results. In Sec.~\ref{SecEasyPlane} we briefly review the easy-plane limit of the XXZ spin chain. In Sec.~\ref{SecExp} we discuss numerical and experimental studies of transport in spin chains. We conclude in Sec.~\ref{SecConclusion} with a discussion of open questions.

\section{Background} \label{SecBackground}

\subsection{Thermal states of integrable systems}\label{scd}

The defining property of integrable systems is that they possess extensively many local conserved charges. A strictly local conserved charge is an operator $\hat Q$ that can be written as a translation-invariant sum of local operators, $\hat Q = \sum_i \hat q_i$ where the operator $\hat q_i$ has support (i.e., acts nontrivially) only on finitely many sites in the thermodynamic limit. (One can also define quasilocal charges~\cite{2016arXiv160300440I} analogously, by allowing $\hat q_i$ to have exponential tails. For our purposes the distinction between these two concepts will not be important.) The stationary states of an integrable system are generalized Gibbs states, which are the highest-entropy states subject to a set of conservation laws. If $\hat Q_n$ are conserved charges that commute with one another, one can represent a generalized Gibbs state as the density matrix $\hat\rho \propto \exp\left(-\sum_n \beta_n \hat Q_n\right)$. 

An equivalent but more physically transparent way of characterizing equilibrium states of integrable systems is in terms of \emph{quasiparticles}. One imagines (in the spirit of the coordinate Bethe ansatz) constructing a general eigenstate on a ring of size $L$ as follows. Begin with a trivial vacuum state (e.g., the $\ket{\downarrow\downarrow\ldots\downarrow\downarrow}$ state in the case of the XXZ spin chain) and start adding particles. In general, once many particles are present, they can exchange energy and momentum through collisions. However, in an integrable system the kinematics of collisions are highly restricted, so that essentially all collisions just involve forward scattering. Thus, each quasiparticle moves ballistically, but in a medium with a refractive index that is determined by the occupation pattern of the other quasiparticles. The eigenstates on a ring are labeled by the quasimomenta (or ``rapidities'') of each quasiparticle, $\{ \lambda_i \}_{i = 1}^N$ for an $N$-particle state. In the free-fermion limit these rapidities would just be the occupation numbers of the single-particle eigenstates. In an interacting integrable system, however, the quantization condition for each quasiparticle depends on the state of the rest of the system, so one computes the set of rapidities $\{ \lambda_i \}$ in an eigenstate by solving the coupled Bethe equations.

Note that the integrable lattice models of interest here host not only elementary quasiparticles, but also bound states of quasiparticles; these bound states are referred to as Bethe strings, and have been experimentally observed~\cite{StringExp}. Because of integrability, strings are \emph{strictly stable} at arbitrary energy density. Therefore, we can treat each string type as a separate species of quasiparticle, indexed by the label $s$. 

We will be interested in characterizing states in the thermodynamic limit, where the allowed quasiparticle labels form a continuum. To this end we define the \emph{density} of quasiparticles of species $s$ in a narrow rapidity window around $\lambda$ as $\rho_s(\lambda)$. One can analogously define a density of states  $\rho^{\mathrm{tot}}_s(\lambda)$: as discussed above, each quasiparticle affects the quantization condition for all the others, so the density of states must generally be determined self-consistently. It is useful to define an ``occupation factor'' $n_s(\lambda) \equiv \rho_s(\lambda)/\rho^{\mathrm{tot}}_s(\lambda)$; again, in a free system in equilibrium this reduces to the Fermi function, but in general it has to be evaluated by solving a set of coupled equations. The thermodynamic Bethe ansatz (TBA) formalism~\cite{Takahashi} provides a straightforward procedure for computing these quantities. 

Thus there are two ways of characterizing stationary states of integrable systems: either in terms of conserved charges or in terms of quasiparticle distribution functions. These descriptions are related through a mapping called ``string-charge duality''~\cite{1742-5468-2016-6-063101}; for our purposes, the key point is that the two descriptions contain the same information. In what follows we will primarily characterize equilibrium states using the quasiparticle distribution function, as the GHD equations have a more intuitive formulation in this picture. 

\subsection{Generalized hydrodynamics}

\subsubsection{Kubo formula}

The simplest dynamical properties of a many-body system to calculate are linear-response transport coefficients like the optical conductivity $\sigma(\omega)$, i.e., the current $J$ generated in response to a sufficiently small external electric field. For any charge $\hat Q$, the current $\hat J = \sum_x \hat j(x)$ is defined by the continuity equation $\partial_t \hat q(x) + \partial_x \hat j(x) = 0$. Here, the derivative is to be interpreted as a discrete derivative in the usual sense. The linear-response conductivity about an equilibrium state can be expressed in terms of a Kubo formula. Here we will be concerned with the high-temperature limit, where the Kubo formula takes the following simple form:
\begin{equation}
T \sigma(\omega) = \frac{1}{L} \int_0^{\infty} dt \langle \hat J(t) \hat J(0) \rangle e^{i \omega t}.
\end{equation}
Note that although $\sigma(\omega) \to 0$ in the high-temperature limit (as the system is already in the maximum entropy state and cannot respond to any perturbations) the quantity $T\sigma(\omega)$ has a finite limit. The quantity $T\sigma(\omega)$ is related to the density-density correlator $C(x,t) = \langle \hat q(x,t) \hat q(0,0) \rangle$ via the continuity equation. This leads to a generalized Einstein relation between the variance of the structure factor, and the Kubo current-current correlator (see {\it e.g.} Refs~\cite{dbd2,2022arXiv220508542S}):
\begin{equation}
\frac{1}{2}\frac{d^2}{d t^2}\int dx x^2 C(x,t) = \frac{1}{L}  \langle \hat J(t) \hat J(0) \rangle .
\end{equation}
For a diffusive system, the variance of the dynamical structure factor goes as $2 \chi D t$ with $D$ the diffusion constant and $\chi = T^{-1} \int dx C(x,t)$ the susceptibility, and we recover the usual Einstein relation $\sigma^{\rm d.c.} = \chi D$ with the d.c.~conductivity $\sigma^{\rm d.c.} = \frac{1}{L} \int_0^{\infty} dt \langle \hat J(t) \hat J(0) \rangle $. 

The dynamical scaling exponent $z$ can be defined by the relation 
\begin{equation}
\int dx x^2 C(x,t) \sim t^{2/z}.
\end{equation}
Using this and the relationship between $C(x,t)$ and $\sigma(\omega)$ we see that $\sigma(\omega) \sim \omega^{1 - 2/z}$. In the limit $z \to 1^+$ the conductivity becomes increasingly peaked around $\omega = 0$. When ballistic transport is present, $\sigma(\omega) = \pi D_{\rm Drude} \delta(\omega) + \ldots$, where the coefficient of the $\delta$-function is called the Drude weight. This result is intuitive: ballistic transport means injected currents do not entirely decay, and the Drude weight is precisely the non-decaying fraction.

Computing $\sigma(\omega)$ reduces to computing dynamical correlation functions of the form $\langle \hat J(t) \hat J(0) \rangle$. Since the Bethe ansatz gives us (in principle) access to all the eigenstates of the many-body system, one might try to evaluate this expectation value by writing it in the Lehmann form
\begin{equation}
\int_{-\infty}^{\infty} dt e^{i\omega t} \langle \hat J(t) \hat J(0) \rangle = \sum_{mn} |\langle m |\hat J | n \rangle|^2 \delta(\omega - (E_m - E_n)),
\end{equation}
where $\ket{m,n}$ are many-body eigenstates with energies $E_m, E_n$. Unfortunately, the task of writing down matrix elements of $\hat J$ (also known as ``form factors'') between pairs of highly excited many-body eigenstates in a way that allows one to take the thermodynamic limit has proved extremely challenging. 


\subsubsection{Mazur bounds} 

Although directly evaluating the Kubo formula is challenging, we are most interested in its $t \to \infty$ limit, as this is where the response is potentially singular. In this limit, one can use the following strategy to lower-bound $\mathrm{Tr}( \hat J(t) \hat J(0) ) $. We note that this quantity has the form of an inner product in the operator Hilbert space. Suppose we are able to construct some family $\{ \hat Q_i \}_{i = 1}^m$ of conserved charges. Assume without loss of generality that the $\hat Q_i$ are orthonormal. Then we can expand $\hat J = \sum_{i=1}^m \mathrm{Tr}(\hat J \hat Q_i) \hat Q_i + \hat J_\perp$, where $\hat J_\perp$ consists of (potentially) non-conserved components of the current. Now $\hat J(t) = \mathrm{Tr}(\hat J \hat Q_i) \hat Q_i + \hat J_\perp(t)$. Without making any assumptions about $\hat J_\perp$ this reasoning gives us the lower bound~\cite{PhysRevLett.106.217206}
\begin{equation}
\lim_{t \to \infty} \left\{ \mathrm{Tr}( \hat J(t) \hat J(0) ) \right\} \geq \sum_{i = 1}^m |\mathrm{Tr}(\hat J \hat Q_i)|^2,
\end{equation}
sometimes called a Mazur bound. This lower bound is saturated in cases where we have identified all the conserved quantities, but it is a rigorous lower bound even for a partial set of conserved quantities. Indeed, finding a single conserved charge that has a nontrivial overlap with the current is sufficient for establishing ballistic transport. Conversely, the absence of ballistic transport means the current has no overlap with \emph{any} conserved charges---a feature that would be unnatural unless there were a symmetry-based reason for it.


\subsubsection{GHD and transport without form factors}

We remarked in Sec.~\ref{scd} that it is mathematically equivalent, but often more fruitful, to think of the states of integrable systems in terms of the quasiparticle distribution functions rather than conserved charges. As we now discuss, this perspective allows us to formulate the question of transport in integrable systems without directly tackling form factors.

As usual with hydrodynamics, the first step is to break the system up into mesoscale cells. Each cell is assumed to be locally in a GGE, which can be specified either in terms of the charges in that cell or its quasiparticle distribution function.
We have two tasks: (i)~time-evolving a general quasiparticle distribution, and (ii)~relating the quasiparticle distribution to charge transport. We discuss both of these at a skeletal level, since the details are amply discussed elsewhere and are not relevant to our main themes~\cite{doyon2019lecture}. Conceptually, (i)~is straightforward: each species of quasiparticle propagates ballistically with an effective velocity set by its environment. This gives rise to a collisionless but nonlinear Boltzmann equation of the form
\begin{equation} \label{eqGHD}
\partial_t \rho_s(\lambda)(x,t) + \partial_x (v_s^{\mathrm{eff}}(\lambda)[\vec{\rho}(x,t)]  \rho_s(\lambda)(x,t)) = 0.
\end{equation}
The square brackets here indicate that the effective velocity is a \emph{functional} of the full distribution function in that cell. Thus, in an inhomogeneous state, Eq.~\eqref{eqGHD} is strongly nonlinear. One might worry that a nonlinear equation of this type might develop shocks, but it turns out that the particular structure of this infinite set of equations prevents shocks from occurring~\cite{1751-8121-50-43-435203,DOYON2018570} -- note that in any case, shocks would be melted by diffusive corrections to this Euler-scale equation~\cite{dbd1,ghkv,dbd2}.

This setup can directly be used to compute charge transport. We start with all the cells in the same equilibrium state but disconnected from one another. Next, we create a slight excess of charge in one cell (which changes its quasiparticle distribution), and connect the cells. We evolve the quasiparticle distribution as above, and then use the final quasiparticle distribution in each cell to read off the value of all conserved charges in that cell. This procedure allows us to compute the full dynamic structure factor without direct reference to any matrix elements. 

\subsubsection{Dressed properties of quasiparticles}

Implementing the GHD prescription for $C_{ab}(x,t)$ above a spatially homogeneous reference state yields the intuitively appealing expression:
\begin{equation}
C_{ab}(x,t) = \sum_s \int d\lambda \rho_s(\lambda) (1-n_s(\lambda)) (m_a^{\mathrm{dr}})(s,\lambda)(m_b^{\mathrm{dr}})(s,\lambda) \delta(x - v^{\mathrm{eff}}_s(\lambda) t). 
\end{equation}
Thus each quasiparticle carries some net charge ballistically with its effective velocity. Computing the correlation function thus reduces to computing the dressed properties of a single quasiparticle above a nontrivial background state. 

It remains to describe how the dressed quantities $m^{\mathrm{dr}}, v^{\mathrm{eff}}$ are computed. We outline the basic logic here and refer to Ref.~\cite{doyon2019lecture} for details -- we will also compute these quantities in a simple model in the next section. The dressed charge of a quasiparticle is computed as follows. Suppose we want to compute $m^{\mathrm{dr}}_a(s, \lambda)$. We perturb the equilibrium state with a field $h_a$ that couples to $Q_a$. This leads to a change in the occupation $n_s(\lambda)$. Since the quasiparticles are fermionic, one can write the occupation number as $n_s(\lambda) = 1/(1 + e^{E_0 + h m^{\mathrm{dr}}})$, from which one can read off the dressed charge. Intuitively the dressed charge is the extent to which the quasiparticle occupation is susceptible to the corresponding field. 

The effective velocity is simplest to understand in the ``flea-gas'' picture of GHD~\cite{solitongases}. The propagation of a quasiparticle consists of two parts: free propagation between collisions, and time delays (or jumps) during collisions. Each collision between quasiparticles of type $(s\lambda), (s'\lambda')$ imparts a shift $d(v,w)$. To find $v^{\mathrm{eff}}_s(\lambda)$ one adds up the shifts due to collisions involving that quasiparticle in time $t$, due to other quasiparticles moving with effective velocities $v^{\mathrm{eff}}_{s'}(\lambda')$. This gives rise to an integral equation
\begin{equation}
v^{\mathrm{eff}}_s(\lambda) = v^0_s(\lambda) + \sum_{s'} \int d\lambda' \rho_{s'}(\lambda') d(s,\lambda;s',\lambda') [v^{\mathrm{eff}}_s(\lambda) - v^{\mathrm{eff}}_{s'}(\lambda')].
\end{equation}
The scattering shifts are specific to the model and can be related to the Bethe ansatz data as detailed in Ref.~\cite{solitongases}.





\subsection{Orders of limits; the GHD regime in finite systems}\label{limits}

We close this section with a brief discussion of the regime of validity of GHD. This discussion will be particularly important when we turn to interpreting numerical results below. GHD is strictly valid in the limit of late times and large systems; however, the thermodynamic limit is always assumed to be taken \emph{before} the late-time limit. Let us take the system size to be $L$, the time to be $t$, and the Lieb-Robinson velocity~\cite{Lieb2004} to be $v_{\mathrm{LR}}$. Then GHD governs the behavior of correlation functions in the limit
\begin{equation}
\lim_{t \to \infty} \left( \lim_{L \to \infty} \langle \hat O(x, t) \hat O(0,0) \rangle_L \right), \qquad x/(v_{\mathrm{LR}} t) \to 0.
\end{equation}
The thermodynamic limit and late-time limit \emph{do not} generally commute. This can be seen in the quasiparticle picture: in the GHD regime, a quasiparticle moves parametrically farther than it spreads. This is not true at very late times in a finite system, since it cannot move farther than $L$ but can delocalize over the entire system. 

Numerical and experimental studies are, of course, limited to finite times and system sizes. To remain in the regime where GHD can reasonably be applied, we must work at times $t \leq L/v_{\mathrm{LR}}$, i.e., times at which quasiparticles have not yet wrapped around the system. This restriction must be kept in mind especially when interpreting finite-size numerics on observables like the conductivity: GHD has nothing to say about this quantity at frequencies  $\omega < v_{\mathrm{LR}} / L$. There is numerical evidence for transport anomalies in this regime~\cite{PhysRevB.104.115163, mierzejewski2022multiple} (especially when integrability is weakly broken in a finite system~\cite{PhysRevB.105.214308}); explaining these observations remains an interesting open question.

\subsection{The space of relevant models}

Our discussion will focus on the spin-1/2 XXZ spin chain, which is a canonical model of quantum magnetism. In conventional (``analog'') experiments this is the most natural model to realize. However, in digital quantum simulators that implement quantum circuits, or in classical tensor-network simulations, it is often helpful to consider ``Trotterized'' models with discrete time evolution. In addition, since much of the long-wavelength physics we will discuss is common to quantum and classical integrable systems, scalable numerical tests of some of the phenomena we discuss are easiest to perform for classical dynamics. The most obvious discrete-time or classical versions of the XXZ model are not integrable; however, integrable Trotterizations and classical limits have been constructed, and we will briefly introduce them in the rest of this section. 

\subsubsection{Integrable Trotterization}

Many numerical approaches (such as the TEBD algorithm~\cite{PhysRevLett.91.147902,PhysRevLett.93.040502}) as well as experiments on noisy quantum computers rely on discretizing quantum evolution. The standard method for doing this is the Trotter decomposition~\cite{PhysRevLett.93.040502}, which consists of breaking up the time evolution $\exp(-i \hat H t) \approx \exp(-i \hat H_{\mathrm{even}} t) \exp(-i \hat H_{\mathrm{odd}} t)$ where $\hat H_{\mathrm{even/odd}}$ consist of terms on even (odd) bonds. Since $\hat H_{\mathrm{even}}$ is a sum of mutually commuting terms, one can write $\exp(-i \hat H_{\mathrm{even}} t)$ as a product of two-site unitary operations on even bonds, which can be implemented simultaneously (and likewise with $\hat H_{\mathrm{odd}}$). Thus the Trotter decomposition expresses $\hat H$ as a brickwork unitary circuit. 
However, it introduces an error at $O(t^2)$, and breaks integrability at the same scale. To preserve integrability, one is forced to use a short timestep (and thus implement many layers of gates, which is potentially demanding for numerical algorithms and especially for small-scale quantum computers). 

Ref.~\cite{PhysRevLett.121.030606} provided an ingenious solution to this issue by writing down a discrete-time family of brickwork unitary circuits that are integrable \emph{for arbitrary time step} and that reduce to the XXZ spin chain in the limit of small time step. Both numerical and analytical results suggest that the Trotterized XXZ spin chain has qualitatively the same transport properties as the Hamiltonian XXZ spin chain~\cite{PhysRevLett.122.150605}. To explore these shared transport properties, it therefore suffices to simulate the integrable Trotterization of the XXZ spin chain at finite time step. For a fixed number of gates, the Trotterized evolution allows one to get to later times, relative to the system's dynamical timescales. 

\subsubsection{Classical integrable spin chains}

The naive classical limit of the XXZ spin chain consists of taking the spin $s$ from $1/2$ to $\infty$. The resulting model is not integrable. However, the closely related Ishimori model 
\begin{equation}
{\cal H} = -\ln \left( 1 + \sum\nolimits_i \mathbf{S}_i \mathbf{\cdot S}_{i+1} \right),
\end{equation}
is integrable~\cite{Ishimori1982}. As we will see below, there is a precise sense in which the long-wavelength dynamics of the quantum Heisenberg and classical Ishimori models coincide~\cite{PhysRevLett.125.070601}. Therefore, for extracting long-wavelength transport properties, classical simulation of the Ishimori model has been a valuable guide. 

The standard way of implementing time-evolution under classical Hamiltonians is to discretize them using a symplectic scheme (which is guaranteed to conserve energy). Generic symplectic integrators maintain integrability only up to an error that scales polynomially in the integration time-step. However, one can define an \emph{exact} discrete-time integrable classical dynamics, consisting of symplectic two-spin maps arranged in a brickwork lattice~\cite{1909.03799}. Symplectic maps corresponding to both the isotropic Heisenberg spin chain and its anisotropic deformations have been defined and studied in Refs.~\cite{2003.05957, krajnik2021anisotropic}. 

\subsubsection{Folded XXZ automaton}

In the large-$\Delta$ limit, the quantum dynamics of the XXZ model reduces to a constrained hopping problem, in which up-spins can hop provided that the hop does not change the number of domain walls. This constrained hopping problem can also be realized as a kinetically constrained classical cellular automaton, as detailed in Refs.~\cite{pozsgay2021yang, PhysRevE.104.054123}. An appealing feature of cellular automata is that, unlike classical Hamiltonian systems, they act on the same Hilbert space as the fully quantum problem: they can be regarded as a special class of unitary transformations that map each product state \emph{in the $z$ basis} to another product state in that basis. Thus, for example, questions about eigenstate and operator entanglement, or the dynamics of nontrivial initial quantum states, can be posed in these models~\cite{prosen2016ffa, gopalakrishnan2018facilitated, sg_ffa, ghkv, PhysRevLett.122.250603, buvca2021rule}.

\section{A simple limit: the folded XXZ spin chain}
\label{SecFoldedXXZ}
\subsection{Folded XXZ model}

The usual nature of spin transport in the XXZ spin chain is best understood in the limit of large anisotropy, as $\Delta \to \infty$, the dynamics become constrained to conserve $\sum_i \hat S^z_i \hat S^z_{i+1}$ which counts the number of domain walls in the system. (Here, a domain wall exists between the spins at sites j and j + 1 if they are anti-aligned.) Using standard strong coupling (Schrieffer-Wolff) expansions, the leading effective Hamiltonian in the limit of large anisotropy is simply an XX flip-flop interaction constrained to act on a subspace with fixed domain wall number: 
\begin{equation} \label{eqFolded}
\hat H_{\rm folded} = J \sum_i \frac{1 + 4 \hat S^z_{i-1} \hat S^z_{i+2}}{2} \left( \hat S^x_{i} \hat S^x_{i+1}+ \hat S^y_{i} \hat S^y_{i+1}\right).
\end{equation}
This model is known as the ``folded'' XXZ spin chain~\cite{Folded1,zadnik2021folded}, and the prefactor $\frac{1 + \hat S^z_{i-1} \hat S^z_{i+2}}{2}$ is a projector that ensures that a flip-flop exchange between sites $i$ and $i+1$ can only occurs the spins on sites $i-1$ and $i+1$ are aligned. Without this constraint, this Hamiltonian would simply mapped to free fermions using a Jordan-Wigner transformation. However, we will see that the constraint leads to very non-trivial transport properties that are generic to the XXZ spin chain in the easy-axis $\Delta>1$ regime, including diffusive spin transport co-existing with ballistic energy transport. 

\subsection{Absence of ballistic spin transport}

In addition to spin and domain wall number conservation,
the allowed moves constrain the motion of the domain walls themselves, so that isolated domain walls are frozen, but adjacent pairs of domain walls (corresponding to isolated magnons in some background) can hop. This means that there exist an exponential number of completely frozen states that have no adjacent domain walls. Configurations in which spins up or down always appear in strings of length two or more are completely frozen, and thus the these sectors are dynamically trivial.

For example, the state
$$
\ket{\hole \hole \hole \hole \hole \hole  \particle \particle \particle  \particle \particle \hole \hole \hole \particle  \particle \particle \particle},
$$ 
is completely frozen, that it to say it is an exact eigenstate of eq.~(\ref{eqFolded}) with zero eigenvalue. On the other hand, the state 
$$
\ket{\hole \hole {\color{red}\particle} \hole \hole \hole \particle \particle \particle \particle \particle \hole \hole \hole \particle \particle \particle \particle}
$$
has a single spin flip (magnon) excitation highlighted in red in the leftmost domain. Upon acting on this state with the Hamiltonian, the magnon, initially a spin up in a majority of spins down, can move around and become a spin down in a majority of spins up. To see this, note that this magnon can hop to the left or right until it collides with a frozen string of up spins.  
For example, after two hops of the magnon to the right, the resulting configuration is
$$
\ket{\hole \hole \hole \hole{\color{red}\particle} \hole  \particle \particle \particle \particle \particle \hole \hole \hole \particle \particle \particle \particle}.
$$
At this point, the up spin can no longer hop to the right, but the down spin on the site immediately to its right \emph{can} hop, resulting in 
$$
\ket{\hole \hole \hole \hole \particle \particle {\color{red}\hole} \particle  \particle \particle  \particle \hole \hole \hole \particle \particle \particle \particle}.
$$
Now, this down spin is the only degree of freedom that can move:  the magnon (i.e., mobile excitation) has changed its charge from $+1$ to $-1$. The magnon can still move to the right until it hits the right boundary of the domain, at which point the configuration again resembles a free particle-like spin up to the right of a frozen up spins domain,
$$
\ket{\hole \hole \hole \hole \particle \particle \particle \particle  \particle \hole {\color{red} \particle} \hole \hole \hole \particle  \particle  \particle \particle}~.~~
$$
This entire sequence of moves can be viewed as a single, unimpeded, magnon-like excitation travelling as a spin up through strings of down spins, and as a spin down through strings of up spins. Each configuration in such single-magnon sectors is connected to exactly two other states by the Hamiltonian, in which the magnon has hopped to the left/right of its current position. The magnon interacts with the otherwise-frozen background domains: As the magnon first hops through a domain from left to right, the position of the domain itself shifts two sites to the left. As a result, the single magnon sector contains $\sim L^2$ states for a system of size $L$, since the magnon has to go around the system $\sim L/2$ times to go back to its original configuration. Despite this unusual feature, the magnon moves like a free object with dispersion relation $\epsilon = - \cos k$ and velocity $v_k = \partial_k \epsilon_k = \sin k$, with the smallest non-zero ``momentum'' $k \sim L^{-2} $ which corresponds to the translations of the magnon~\cite{2021arXiv210802205S,KnapTracer}.

\begin{figure}[tb]
\begin{center}
\includegraphics[width = 0.4\textwidth]{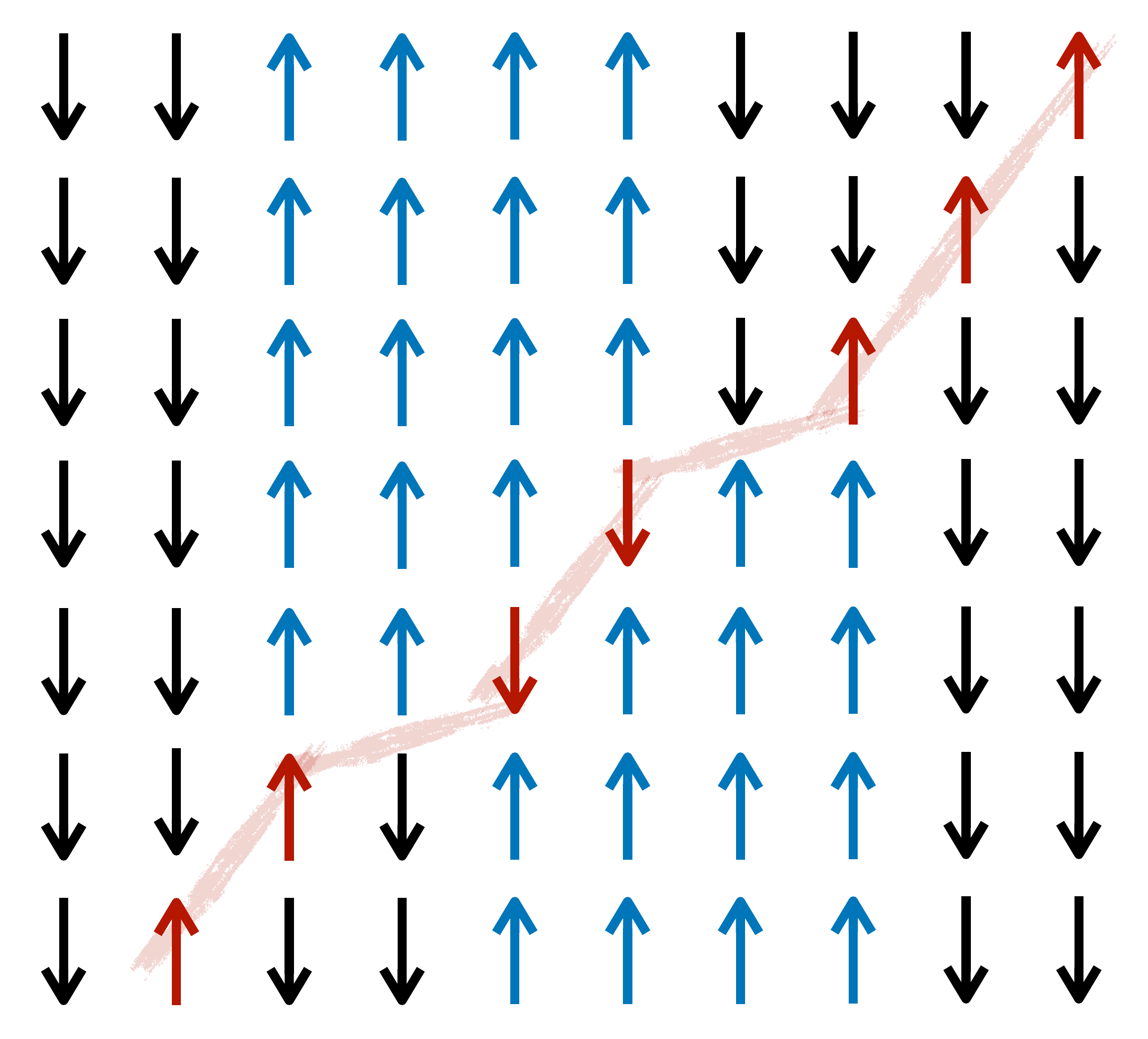}
\caption{{\bf Magnon propagation in the folded XXZ model.} Example of magnon propagation: initially a spin up in a majority of spins down, can move around and become a spin down in a majority of spins up. Note the displacement (or Wigner time delay) of the quasiparticles after collision. }
\label{figMagnon}
\end{center}
\end{figure}

Since the mobile magnons have a free dispersion, and move ballistically,  energy transport is ballistic. The corresponding delta peak in the thermal conductivity is characterized by the thermal Drude weight $D^E_{\rm Drude} = \int \frac{dk}{2 \pi} n_k (1-n_k) v_k^2 \epsilon_k^2$, where $n_k$ is the occupation number (Fermi factor) of the magnons. At infinite temperature, we have $n_k = \frac{1}{4}$ since magnons correspond to local configurations $\downarrow \uparrow \downarrow$ or $\uparrow \downarrow \uparrow$, so two out of eight 3-spin configurations. Correspondingly, the energy-energy dynamical correlation function is given, at the Euler scale
(that is, at the scale $x \sim t$ and ignoring {\it e.g.} diffusive corrections) by 
\begin{equation}
\langle \hat \epsilon(x,t) \hat \epsilon(0,0) \rangle = \int \frac{dk}{2 \pi} n_k (1-n_k) v_k^2 \epsilon_k^2 \delta(x- v_k t).
\end{equation}
Naively, as the magnons move ballistically and carry spin one, one would similarly expect ballistic spin transport. However, as we discussed above, as it moves through domains of up and down spins, its magnetization oscillates between $\pm 1$ (Fig.~\ref{figMagnon}). As a result, its magnetization  is ``dressed'' to be zero on average $m^{\rm dr}_k=0$. The spin Drude weight vanishes~\cite{PhysRevLett.119.020602}
\begin{equation}
D_{\rm Drude} = \int \frac{dk}{2 \pi} n_k (1-n_k) v_k^2 (m^{\rm dr}_k)^2 = 0,
\end{equation}
indicating sub-ballistic spin transport. 

\subsection{Origin of diffusive spin transport}

\subsubsection{Depolarizing quasiparticle perspective.}

In order to uncover the nature of spin transport in the folded XXZ model, we need to reintroduce dynamics and time in the ``depolarizing'' process of the magnon leading to $m^{\rm dr}_k=0$, following Ref.~\cite{GV19}. By central limiting arguments, the region through which the magnon propagates ballistically on a
timescale $t$ has ${\cal O}( \sqrt{|v_k t|})$ more $\uparrow$ domains than $\downarrow$ domains. As a result, we expect that the ``dynamical'' dressed magnetization to go as $m^{\rm dr}_k \sim \frac{1}{\sqrt{|v_k t|}}$. Magnons thus move ballistically but with vanishing magnetization $t^{-1/2}$, so that over a timescale $t$ magnetization spread over a distance $\sqrt{t}$~\cite{GV19}, corresponding to diffusing scaling. Note that the origin of diffusion here is very different from that of the usual diffusive behavior expected in chaotic system, which comes from integrating out fast degrees of freedom. This will have important consequences for quantities such as full counting statistics that we will explore below. 

In order to make this argument for diffusion more precise, we focus on the spin dynamical structure factor
\begin{equation}
C(x,t) = \langle \hat S^z (x,t) \hat S^z(0,0)\rangle, 
\end{equation}
at infinite temperature. If transport were ballistic, the spatial variance of the structure factor would scale as $t^2$, with a weight given by the spin Drude weight
\begin{equation}
\int dx x^2 C(x,t) = D_{\rm Drude} t^2 + \dots = t^2 n (1-n) \int \frac{dk}{2 \pi} v_k^2  (m^{\rm dr}_k)^2  + \dots
\end{equation}
Here we have used $n_k = n = \frac{1}{4}$ independent of momentum for the magnon at infinite temperature. The dressed magnetization is understood as a dynamical quantity, with $m^{\rm dr}_k \sim 1/\sqrt{t}$ as explained above, so that the variance of the structure factor scales linearly with time, corresponding to diffusion. The prefactors are harder to obtain purely intuitively, but using thermodynamic Bethe ansatz (TBA) results, one finds that the variance of the dressed magnetization over an interval of size $\ell = |v_k t|$
\begin{equation}
\langle (m^{\rm dr}_k)^2 \rangle= \frac{16}{9} \frac{1}{\ell}.
\end{equation}
Using this formula with $n=1/4$, we find 
\begin{equation}
 \int dx x^2 C(x,t) =  \frac{t}{3} \int_{-\pi}^\pi \frac{dk}{2 \pi} |v_k| + \dots = \frac{2 t}{3 \pi}  + \dots 
\end{equation}
This corresponds to a diffusive variance $2 \chi D t$ of the structure factor, with susceptibility $\int dx C(x,t) = 1/4$ and
diffusion constant~\cite{GV19, dbd2,KnapTracer} 
\begin{equation} \label{eqXXZfoldedDiffusionConstant}
D= \frac{4}{3 \pi}.
\end{equation}
Despite the unusual nature of this diffusion process, the dynamical structure factor has a diffusive form 
\begin{equation}
C(x,t) = \frac{1}{8 \sqrt{ \pi D t}} {\rm e}^{-\frac{x^2}{4 D t}}.
\end{equation}

\subsubsection{Moving domain-wall perspective.}
There is a ``dual'' way to understand diffusion, which focuses on the motion of the frozen domains instead of the depolarizing magnons~\cite{dbd1,dbd2,GV19}. In the absence of magnons, the domain walls are all frozen: $v_{\rm DW} =0$, and the background pattern remains immobile $x_{\rm DW} = v_{\rm DW} t =0$. However, the position and velocity of the domain walls fluctuate because of the magnons colliding with it. Using the chain rule, the variance of the position of the background can be expressed as~\cite{ghkv}
\begin{equation}
\delta x_{\rm DW} ^2 = t^2 \int \frac{d k }{2\pi} \int \frac{d q }{2\pi} \frac{\delta v_{\rm DW}}{\delta n_k}  \frac{\delta v_{\rm DW}}{\delta n_q} \langle \delta n_k \delta n_q \rangle,
\end{equation}
where the equilibrium fluctuations of the Fermi factors of the magnons over an interval of length $\ell$ are given by $ \langle \delta n_k \delta n_q \rangle = 2 \pi \delta(k-q) \frac{n_k (1-n_k)}{\ell}$. Here we choose $\ell = |v_k| t$, which is the distance over which magnons with pseudo-momentum $k$ can move over time $t$ and collide the domain wall which on average has zero velocity. The functional derivative of the velocity of the domain wall (background) with respect to the Fermi factors of the magnons is given by generalized hydrodynamics~\cite{ghkv}
\begin{equation}
\frac{\delta v_{\rm DW}}{\delta n_k} = |v_k| \Delta x_{\rm DW},
\end{equation}
where $\Delta x_{\rm DW} = \frac{8}{3}$ is the ``dressed'' displacement of the background after a collisions with a magnon -- which is different from the bare displacement $\Delta x_{\rm DW} =2 $. The derivation of those formulas is more technical and relies on TBA and GHD techniques, and we refer the interested readers to Refs.~\cite{ghkv,dbd1,dbd2} for technical details. Using those results, we find
\begin{equation}
D_{\rm DW} = \frac{1}{2} n(1-n) (\Delta x_{\rm DW})^2 \int \frac{d k }{2\pi} |v_k| = \frac{4}{3 \pi},
\end{equation}
in agreement with eq.~(\ref{eqXXZfoldedDiffusionConstant}). Therefore, the spin diffusion constant can either be interpreted as ballistic magnons that progressively get depolarized, or as the ``jiggling'' motion of the frozen background domains as magnons collide with domain walls. This dual interpretation implies a relation between the dressed magnetization of the magnons and the dressed displacement of the domain walls:
\begin{equation}
\frac{1}{2}(\Delta x_{\rm DW})^2 =  \lim_{h \to 0} \partial_h^2 m^2_k,
\end{equation}
which was dubbed ``magic formula'' in Ref.~\cite{NMKI19}.

\subsection{A new class of diffusion: anomalous full counting statistics}

As discussed above, at the level of linear response, spin transport in the folded XXZ model is diffusive. However, the physical mechanism is highly unusual, with interesting consequences going beyond hydrodynamic expectation values. Experiments in platforms such as ultracold atoms or superconducting qubit arrays are not limited to measuring expectation values: quantum measurements lead to simultaneous snapshots of all the particles in a system~\cite{gritsev2006full, hofferberth2008probing, mazurenko2017cold, gross2017quantum, arute2019quantum, scholl2021microwave}. A quantity that compactly encapsulates the statistics of snapshots is the ``full counting statistics'' (FCS) of conserved charges~\cite{levitov1993charge,doi:10.1063/1.531672,PhysRevB.51.4079,PhysRevB.67.085316,PhysRevLett.110.050601, TOUCHETTE20091, garrahan2018aspects, Essler_xxz_fcs, Essler_Ising_fcs}, in our case, spins. We consider the following setup: initialize the folded XXZ spin chain with both the left and right half-systems in definite magnetization $\langle \hat{S}^z \rangle = \pm m/2 $. Then, after evolving for a time $t$, measure the magnetization transfer $\hat{Q}(t)$ to the left half-system. The probabilistic classical outcome is denoted $Q(t)$. Repeating this experiment many times yields a quantum distribution of measurement outcomes $P(Q(t))$. The first cumulant $\langle \hat{Q}(t) \rangle \sim \sqrt{t}$ of the FCS is related to hydrodynamic transport which in our case is diffusive, while the higher cumulants go beyond it. For a typical many-body diffusive system, classical or quantum, we expect the fluctuations of $\hat{Q}(t)$ to be controlled by the variance scaling as $\langle \hat{Q}^2(t) \rangle \sim \sqrt{t}$, corresponding to a central limit behavior $\sqrt{\langle \hat{Q}^2(t) \rangle}/\langle \hat{Q}(t) \rangle \sim \frac{1}{\sqrt{t}} $. However, in the folded XXZ spin chain, since diffusion is due to the motion of a single particle, namely the domain wall at the origin, its FCS is anomalous. This was first observed numerically in Ref.~\cite{PhysRevLett.128.090604}, and through an exact solution of a cellular automaton in Ref.~\cite{krajnik2022exact}. Here we follow~\cite{2022arXiv220309526G}, and use the physical picture outlined above. 

\begin{figure}[tb]
\begin{center}
\includegraphics[width = 0.65\textwidth]{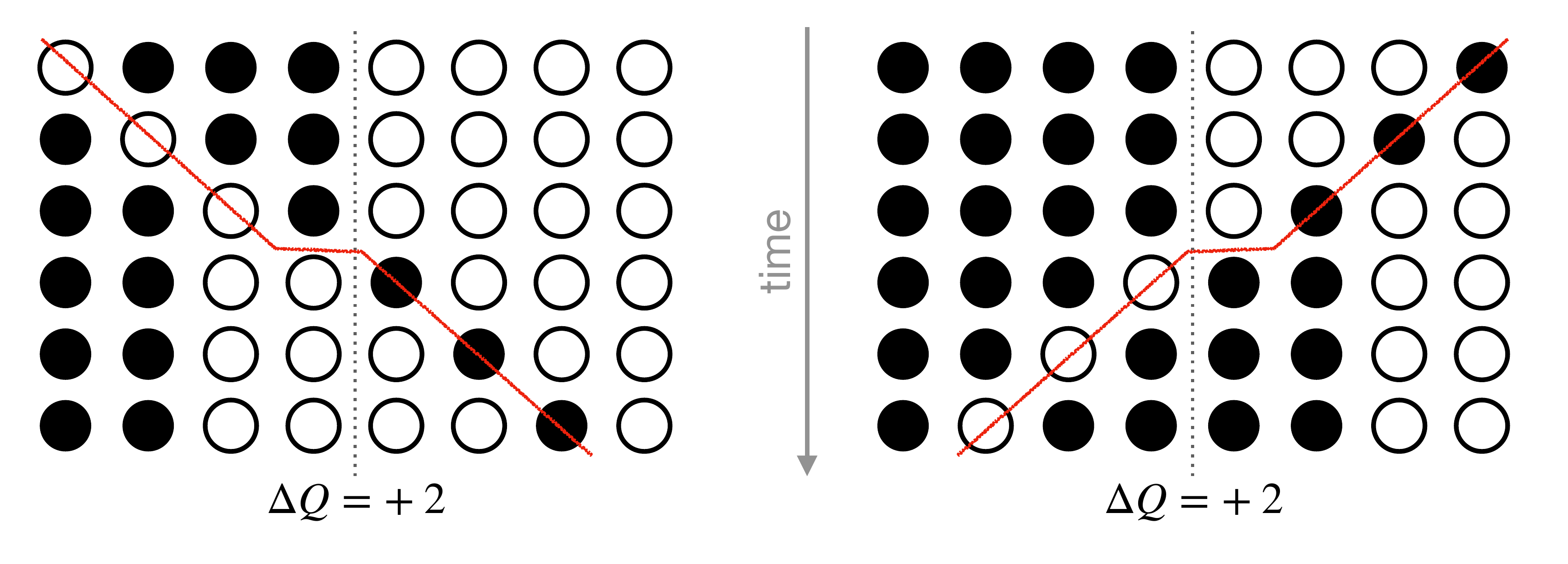}
\caption{{\bf Full counting statistics setup.} Illustrative trajectories in which a magnon goes through the domain wall starting either from the left or the right. Here spins up are represented as full circles, while empty circles correspond to down spins. The magnon trajectory is indicated in red; the domain wall dividing the left and right half-chains is shown as a gray dashed line. The passage of the magnon shifts the domain wall by two steps. Whichever way the domain wall shifts, provided it moves away from the cut, one unit of magnetization (i.e., one net black circle) moves from the left half-chain to the right half-chain. If we run this process in reverse, the domain wall moves toward the cut and magnetization is transferred to the left. Figure reproduced from Ref.~\cite{2022arXiv220309526G}.}
\label{figFCS}
\end{center}
\end{figure}

As magnons collide with the domain wall (DW), initially at the origin in this setup, it undergoes diffusive motion as explained above. In order to relate the the motion of the domain wall to the magnetization transfer $Q(t)$, we make the following key observation in the limit $m \to 1$ (fully polarized DW): whenever the DW moves away from the origin, $Q(t)$ goes up; whenever the DW moves toward the origin, $Q(t)$ goes down (Fig.~\ref{figFCS}). Thus the distribution $P(Q(t))$ in this limit is simply the distribution of the distance of the DW from the origin, i.e., the absolute value of a random walk. Away from this limit, by inspection one finds that the magnetization transfer $Q(t) = |x_{\rm DW} (t)| m_{[0,x_{\rm DW} (t)]}$, where $x_{\rm DW}$ is the position of the DW, and $m_{[0,x_{\rm DW} (t)]}$ is the magnetization density in the interval $[0,x_{\rm DW} (t)]$. The position of the DW is simply a Brownian motion normally distributed with variance $2 D t$ and zero mean (with $D$ the diffusion constant of the DW, which coincides with the spin diffusion constant as $m \to 0$), while using standard statistical mechanics results $m_{[0,x]}$ is a normally distributed variable with mean $m$ and variance $(1 - m^2)/|x|$. Combining those results, we find the distribution of the magnetization transfer~\cite{2022arXiv220309526G}: 
\begin{equation}
P(Q) = \int_0^\infty dx \frac{\exp \left(-\frac{x^2}{4 D t}-\frac{x \left(\frac{Q}{x}-m\right)^2}{2 \left(1-m^2\right)}\right)}{\sqrt{2 \pi^2 (1-m^2)} \sqrt{D t x}}. 
\end{equation}
This distribution is non-Gaussian even in the equilibrium case $m=0$, where it characterizes magnetization transfer at the scale $Q \sim t^{1/4}$. In the presence of a bias $m>0$, at long times the distribution approach a half-Gaussian for the rescaled variable $Q/\sqrt{t}$. As a result, the standard deviation of charge transfer is of the same order as the mean, in sharp contrast with the central limit theorem expectation. The breakdown of the central limit theorem is physically very transparent: magnetization transfer is due to the random walk of a {\em single} particle, namely the DW (frozen background). 

\subsection{Away from integrability: XNOR model} \label{SecXNOR}

The folded XXZ model is clearly integrable, as the number of each string (frozen domains) is conserved by the time evolution. Since spin transport is diffusive already in the integrable limit, one could naively expect that it would remain diffusive upon breaking integrability. However, because of emergent conservation of domain wall number from the limit $\Delta \to 1$, the dynamics remains constrained even in the non-integrable limit. As a result of the constraint, only single magnons are mobile even after breaking integrability, while bigger domains remain immobile. Upon breaking integrability, the motion of the magnons is now diffusive: over a timescale $t$, they move over a distance $\sqrt{t}$, their dressed magnetization is $\sim t^{-1/4}$, so the net magnetization transfer is $\sim t^{1/4}$~\cite{2021arXiv210802205S,2021arXiv210913251D}. Accordingly, the frozen domain now jiggles because it gets hit by diffusive magnons, so its motion is also $\sim t^{1/4}$.

This corresponds to {\em subdiffusion}, with dynamical exponent $z=4$. This is a first example of how proximity to integrability can lead to unexpected anomalous transport properties, distinct from those of the integrable limit. 
In general, chaotic dynamics with such an ``XNOR'' constraint, namely spin flips between sites $i$ and $i+1$ allowed only if neighboring spins on sites $i-1$ and $i+2$ are aligned is indeed subdiffusive with $z=4$: this was shown analytically using constrained quantum circuits with a $U(1)$ charge~\cite{kvh,rpv} in Ref.~\cite{2021arXiv210802205S}. Note that this universality class of subdiffusive transport with $z=4$ is different from the $z=4$ dynamics obtained from fracton-like constraints~\cite{PhysRevLett.125.245303,PhysRevX.10.011042,PhysRevB.100.214301,PhysRevResearch.2.033124} -- in particular, scaling functions are different. 

\section{Easy-axis XXZ model} \label{SecEasyAxis}

We now turn to transport in the easy-axis regime $\Delta >1$ of the XXZ spin chain. Most of the physics is essentially identical to that of the folded XXZ model detailed above. However, generalizing explicit formulas away from the $\Delta \to \infty$ limit requires some nontrivial steps: the magnons are at high density, and magnon strings of all lengths are mobile, and must be accounted for. However, these issues can be addressed at the level of GHD~\cite{Doyon,Fagotti}: we assume that quasiparticles are in local equilibrium in an appropriate generalized Gibbs ensemble, and evaluate the dressed quasiparticle dispersion as well as the quasiparticle distribution function using data from the thermodynamic Bethe ansatz solution. 

\subsection{Quasiparticle content}

To proceed, we will need the following standard thermodynamic Bethe ansatz (TBA) concepts, already summarized in Sec.~\ref{SecBackground}: an equilibrium state is uniquely characterized by a density $\rho_s(\lambda)$ of
quasiparticles with quantum numbers $(s,\lambda)$. Here $s = 1,2 , \dots$ labels ``strings'', where $s=1$ corresponds to magnons, while $s>1$ are magnon bound states, corresponding to the frozen domains of $s$-spins in the folded XXZ limit. The dispersion of those quasiparticles is parameterized by the rapidity $\lambda$.
We also introduce the available density of states $\rho^{\mathrm{tot}}_s(\lambda)$ and
the associated Fermi filling fractions $n_s(\lambda) \equiv \rho_s(\lambda)/\rho^{\mathrm{tot}}_s(\lambda)$. Finally, like in the folded XXZ limit, interactions ``dress'' the group velocity and magnetization (along with other local charges) carried by quasiparticles; we denote these $v^{\mathrm{eff}}_s(\lambda)$ and $m_s^{\mathrm{dr}}(\lambda)$ respectively, as in Sec.~\ref{SecBackground}.

\begin{figure}[tb]
\begin{center}
\includegraphics[width = 0.45\textwidth]{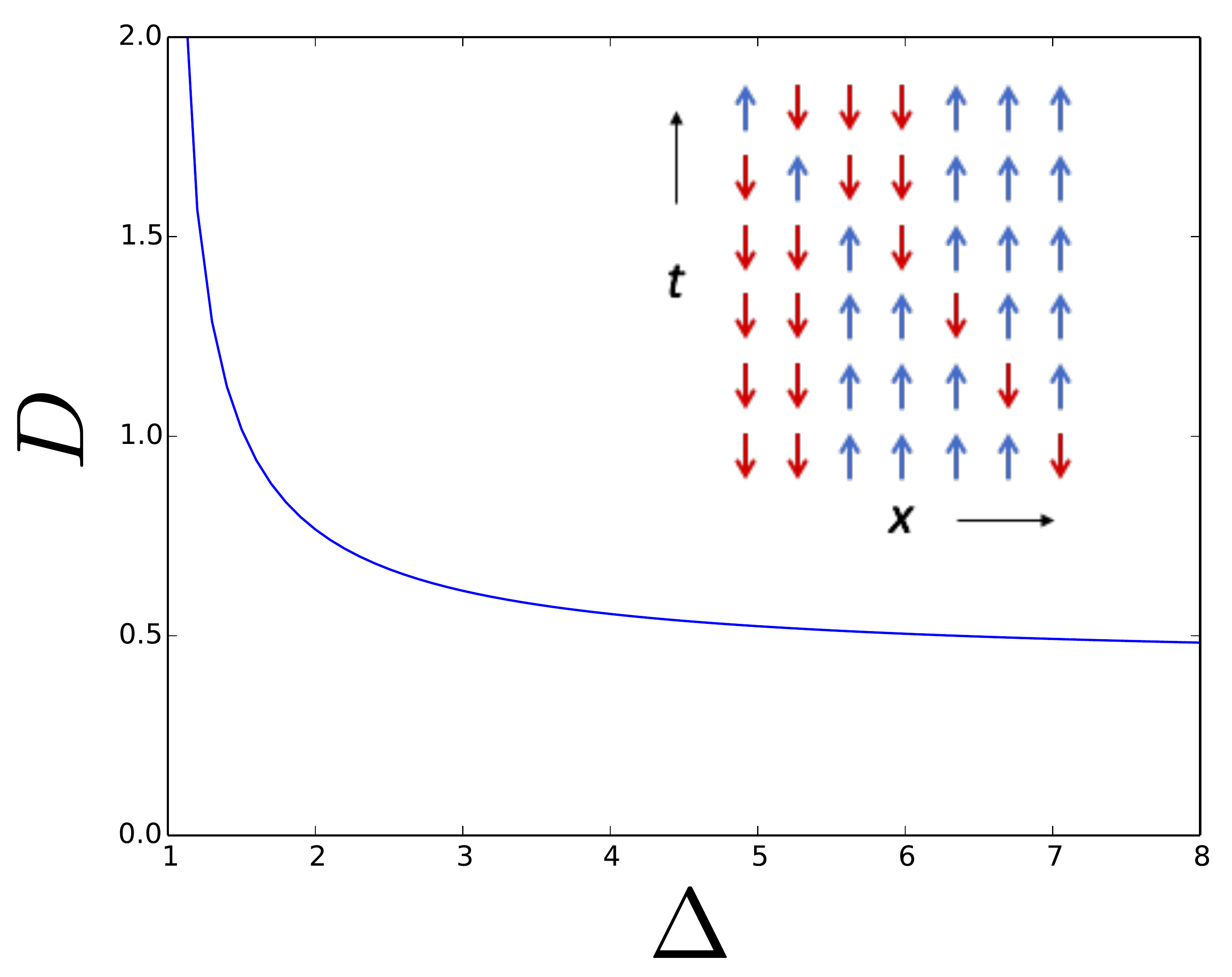}
\caption{{\bf Spin diffusion constant $D$ in the easy-axis phase $\Delta >1$ as a function of $\Delta$}. The diffusion constant diverges in the anisotropic limit, leading to superdiffusion. {\it Inset:} Propagation of a magnon in the large $\left| \Delta \right|$, low temperature, ferromagnetic limit. Reproduced from Fig.~\cite{GV19}.}
\label{FigD}
\end{center}
\end{figure}

\subsection{Diffusion constant}

 We follow the approach of Sec.~\ref{SecFoldedXXZ} generalized to arbitrary $\Delta$ and finite temperature, see Refs.~\cite{GV19,PhysRevLett.125.070601} for the original references. 
In the presence of a small magnetic field $h$, the variance of the spin-structure factor is scaling ballistically with the Drude weight~\cite{GHDII,PhysRevB.96.081118}
\begin{equation}
\sigma^2  \equiv \int dx  \ x^2 \langle \hat S^z(x,t) \hat S^z(0,0)\rangle =  t^2 \sum_{s=1}^\infty \int d \lambda \rho_s(\lambda) (1-n_s(\lambda)) (v^{\rm eff}_s (\lambda)m^{\rm dr}_s(\lambda))^2. 
\end{equation}
As $h\to0$ (or half-filling, in the equivalent fermionic language), the dressed magnetization vanishes, giving rise to normal spin diffusion. To see this, we follow the logic of section~\ref{SecFoldedXXZ} on the folded XXZ chain, and write down the effective dressed magnetization felt by a $(s,\lambda)$-quasiparticle propagating with an effective velocity $v^{\rm eff}_s(\lambda)$,
\begin{equation}
(m^{\rm dr}_s(\lambda))^2 = \frac{1}{2}\left. \partial^2_h (m^{\rm dr}_s)^2 \right|_{h=0} h^2 + \dots,
\end{equation}
where $h^2$ can be though of as the effective field originating from thermal fluctuations:
\begin{equation}
h^2 = \frac{1}{(4 \chi)^2} m^2 = \frac{1}{4 \chi |v^{\rm eff}_s(\lambda) t|}.
\end{equation}
Here we have taken the fluctuations of the magnetization
\begin{equation}
m^2= \frac{4 \chi}{\ell},
\end{equation}
over a distance $\ell=|v^{\rm eff}_s(\lambda) t|$, and the factor of $4$ in front of $\chi$ is to match Bethe Ansatz conventions. This yields
\begin{equation}
\sigma^2 =  t \sum_{s = 1}^\infty \int d\lambda \rho_s(\lambda) (1-n_s(\lambda)) \left| v^{\rm eff}_s(\lambda)\right|  \frac{1}{8 \chi}\left. \partial^2_h (m^{\rm dr}_s)^2\right|_{h=0} . 
\end{equation}
To identity the diffusion constant, we note that the variance of the spin structure factor should scale as $\sigma^2 = \chi 2 D t$. We thus get
\begin{equation}
D  =  \sum_{s = 1}^\infty \int d\lambda \rho_s(\lambda) (1-n_s(\lambda)) \left| v^{\rm eff}_s(\lambda)\right|  \frac{1}{16 \chi^2}\left. \partial^2_h (m^{\rm dr}_s)^2\right|_{h=0} . 
\end{equation}
This simple argument predicts a diffusion constant in agreement with other approaches~\cite{dbd2,PhysRevLett.123.186601}.  This expression for the diffusion constant is valid for $\Delta \geq 1$, and diverges in the isotropic limit $\Delta \to 1$ (Fig.~\ref{FigD}). Plugging in explicit TBA solutions at infinite temperature, we find~\cite{GV19}
\begin{equation}
\label{eqDiffusionConstant}
D = \frac{2 \sinh \eta}{9 \pi} \sum_{s=1}^\infty (1+s) \left[\frac{s+2}{\sinh \eta s} - \frac{s}{\sinh \eta (s+2)} \right],
\end{equation}
with the anisotropy $\Delta=\cosh \eta >1$.

\subsection{Structure factor}\label{secgvw}

 The hydrodynamic expression for the dynamical structure factor takes the form~\cite{GHDII,PhysRevB.96.081118}:
\begin{equation}
C(x,t)   =  \sum_{s = 1}^\infty \!\int d\lambda\, \rho^{\rm tot}_s(\lambda) n_s (1-n_s) (m^{\mathrm{dr}}_s)^2 \varphi_t[x - v^{\rm eff}_s(\lambda) t]. \label{eqGHDSzSz}
\end{equation}
Here, the function $\varphi_t (\zeta)$ is the propagator of a string with quantum numbers $(s, \lambda)$ from $(0,0)$ to $(x,t)$. At the Euler level this propagator would simply be a Dirac delta function, while we take it to be a broadened Gaussian with variance $2 D_{s}(\eta,\lambda) t$ at the diffusive level. The diagonal quasiparticle diffusion constant $D_{s}(\eta,\lambda)$ was computed in Refs.~\cite{dbd1,ghkv}. As discussed above, exactly at half-filling $h\to 0$, all strings but the heaviest ones $s \to \infty$ become effectively neutral as $m^{\rm dr}_s$ goes to 0 for $s \ll h^{-1}$, so the structure factor becomes diffusive 
\begin{equation}
C(x,t)   = \frac{1}{8\sqrt{\pi D(\eta) t}} {\rm e}^{ - \frac{x^2}{4 D(\eta) t}},
\end{equation}
where the diffusion constant $D(\eta)$ is given by eq.~(\ref{eqDiffusionConstant}). 

Away from half-filling (corresponding to a finite field $h>0$), spin transport becomes ballistic as strings with $s \gg h^{-1}$ become magnetized. However, the structure factor still has some interesting local relaxation structure~\cite{gvw}. This arises because the velocity of ``heavy'' (large $s$) strings is strongly suppressed: this can be seen perturbatively at large $\Delta$, as $v^{\rm eff}_s \sim \Delta^{1-s}$, since moving a domain of $s$-spins require going to $s^{\rm th}$ order in perturbation theory. In general, we have $v^{\rm eff}_s \sim {\rm e}^{-\eta s}$ with $\Delta = \cosh \eta$. Therefore, from this ballistic motion strings remain near the origin with amplitude $\sim \frac{1}{v^{\rm eff}_s t} \sim {\rm e}^{\eta s}/t$. Meanwhile, the density of such heavy strings goes as ${\rm e}^{-s h}$, so there are two cases for the structure factor depending on the sign of $\eta -h$. If $h>\eta$, then the contribution from heavy strings is exponentially suppressed in the structure factor, and near the origin, we have $C(x=0, t) \sim t^{-1}$ from light strings, as expected from ballistic scaling. However, if $h<\eta$, then heavy strings dominate the local relaxation of the structure factor. For $s$ very large, the motion of the heavy strings is first diffusive (since their velocity is very small), with a diffusion constant $D$ that is essentially independent of $s$ and rapidity $\lambda$. However, after a time scale defined by $\sqrt{D t} \sim v^{\rm eff}_s t \sim {\rm e}^{-\eta s} t$ their motion is ultimately ballistic. Alternatively this equation can be used to defined a crossover string index $s_*(t) \sim \frac{1}{2 \eta} \log t $ at a given time $t$. The structure factor near the origin is dominated by strings $s<s_*(t)$ which move diffusively and contribute a factor $\frac{1}{\sqrt{D t}}$. In this regime $h< \eta$, we find~\cite{gvw}
\begin{equation}
C(x=0, t) \sim \sum_{s > s_*(t) \sim \frac{1}{2 \eta} \log t} \frac{ {\rm e}^{-h s}}{\sqrt{t}} \sim t^{-1/2 - \frac{h}{2 \eta}}.
\end{equation}
This shows that even if spin transport is ballistic, the structure factor can be highly anomalous, in this case exhibiting continuously varying exponents. A similar mechanism was found in integrable random spin chains in Ref.~\cite{agrawal2019}. 


\begin{figure}
    \centering
   \includegraphics[width = 0.55\columnwidth, , angle =90]{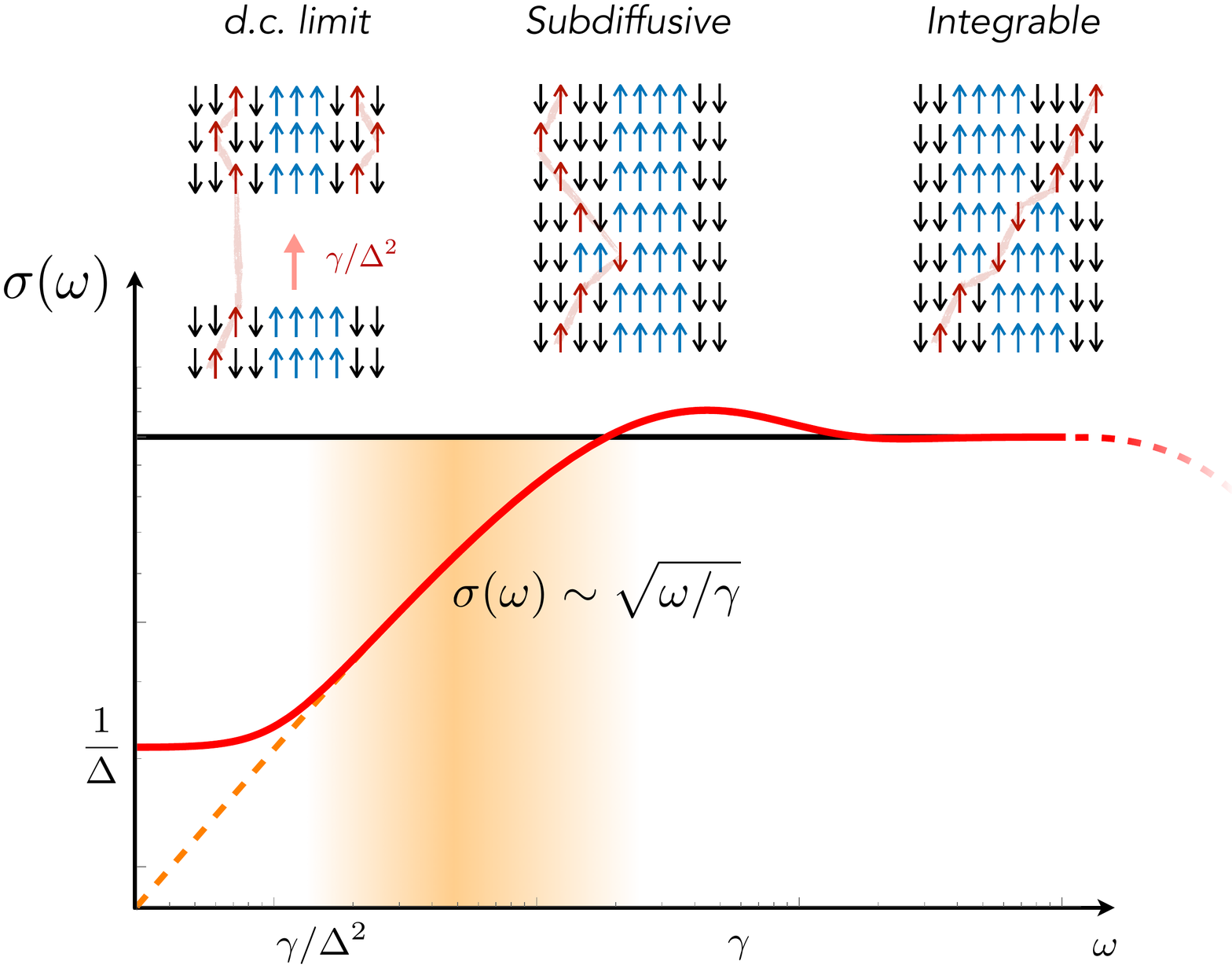}
    \caption{{\bf Anomalous low-frequency spin conductivity in the noisy XXZ chain} (in log-log scale). In the frequency regime $\frac{\gamma}{\Delta^2} \ll \omega \ll \gamma$, spin transport is subdiffusive with $\sigma(\omega) \sim \sqrt{\omega/\gamma}$, corresponding to dynamical exponent $z=4$. At very low frequency, the conductivity eventually saturates to a finite d.c. value proportional to $\Delta^{-1}$. The cartoon on top illustrates the dominant spin dynamical processes at different time-scales: single, mobile, magnons are pictured in red and strings, frozen, in blue. Figure reproduced from Ref.~\cite{2021arXiv210913251D}.  }
    \label{figConductivityXXZlargeDeltaNoise}
\end{figure}

\subsection{Away from integrability: discontinuity of the diffusion constant}

As discussed in Sec.~\ref{SecXNOR}, spin transport at large anisotropy $\Delta \gg 1$ becomes subdiffusive with dynamical exponent $z=4$ upon breaking integrability because of the emergent XNOR constrained dynamics. At long times and finite $\Delta$, we expect to recover diffusion. For concreteness, we focus on uncorrelated noise, and take $\hat H = \hat H_{\mathrm{XXZ}} + \sqrt{\gamma} \sum_i \eta_i(t) \hat S^z_i$, where the noise $\eta$ satisfies $\langle \eta_i \rangle = 0$ and $\langle \eta_i(t) \eta_j(0)\rangle =  \delta(t) \delta_{ij}$. The noise backscatters magnons at a rate $\sim \gamma$: this sets the crossover between the integrable diffusive behavior and $z=4$ subdiffusion. This implies that the conductivity scales as $\sigma(\omega) \sim \sqrt{ \omega/\gamma}$ for $\omega \ll \gamma$.

Noise can also create magnons out of strings, by the following process: one end of a larger string virtually hops away from the rest of the string by one site, with amplitude $1/\Delta$, and is put on shell by the noise, giving a transition rate $\gamma/\Delta^2$. This breaks the constraint, and we expect this process to restore diffusion. This means that subdiffusion $\sigma(\omega) \sim \sqrt{ \omega/\gamma}$ occurs in the broad frequency regime $\gamma \gg \omega \gg \gamma/\Delta^2$ (Fig.~\ref{figConductivityXXZlargeDeltaNoise}). Matching up with the behavior at very low frequencies, we recover diffusion with a conductivity $\sigma_{\rm dc} \sim \frac{1}{\Delta}$~\cite{2021arXiv210913251D}. Remarkably, the conductivity and the diffusion constant are {\em independent} from the noise strength! Moreover, it also means that the diffusion constant in near integrable systems is non-perturbative, and a discontinuous function of the strength of the integrability-breaking couplings.

\section{Heisenberg model} \label{SecXXX}

\subsection{Nature of single quasiparticles}

From the point of view of the Bethe ansatz solution, the isotropic point $\Delta = 1$ is similar to the easy-axis phase $\Delta > 1$ except for one crucial detail: the effective (as well as bare) velocities of the strings. In the easy-axis phase, these velocities fall off exponentially with string size, while at the Heisenberg point they fall off algebraically. This change in the scaling of velocities is linked to a qualitative change in the character of elementary excitations about the symmetry-broken ground state. In the easy-axis phase, these are domains, but at the isotropic point they are Goldstone modes associated with the broken $SU(2)$ symmetry. 

The distinctive transport signatures of the Heisenberg model stem from the bare properties of its quasiparticles, which are best understood as ``Goldstone solitons,'' or wavepackets of Goldstone modes that are stabilized by the nonlinearity of the dynamics. Because the Bethe-ansatz solution begins with the ferromagnetic ground state and constructs a general state by piling on quasiparticles, the relevant Goldstone modes are the quadratically dispersing excitations above an $SU(2)$ ferromagnet~\cite{sachdev2011}. Consider a spatially gradual (but large-amplitude) vacuum rotation above a ferromagnetic ground state: suppose the orientation rotates from the reference ground state by some $O(1)$ amount and then back over a distance $\ell$. The momentum uncertainty of this wavepacket is at least $1/\ell$, so its characteristic group velocity (from the quadratic dispersion) also scales as $v(\ell) \sim 1/\ell$. Again, from the quadratic dispersion, the energy density scales as $1/\ell^2$, and the total energy of the wavepacket scales as $1/\ell$. Finally, since the wavepacket is an $O(1)$ rotation of the vacuum orientation spread out over $\ell$ sites, it carries a total magnetization $\sim \ell$ relative to the ground state. 

We briefly comment on the crossover between the easy-axis and isotropic regimes as $\Delta \to 1^+$. In the easy-axis regime, the (bare) strings are domains of the minority species. The spins at the edges of the domain are energetically confined to a region of size $\eta^{-1} = 1/(\cosh^{-1} (\Delta))$ when the domain is of size $\gg \eta^{-1}$. Quasiparticles smaller than this size cannot resolve the easy-axis anisotropy and behave as they would at the isotropic point. At the isotropic point, the only length-scale associated with a quasiparticle is its own size. 

\subsection{Elementary argument for superdiffusion}

The dressing of the quasiparticle magnetization at the isotropic point is identical to that in the easy-axis phase, for the same physical reason: the smaller quasiparticle encounters the larger one as a local vacuum rotation, and to keep propagating in the rotated vacuum it needs to reorient itself. Therefore, the dressed charge of an $s$-string is zero at half filling for any finite $s$. 

We now discuss how these observations immediately lead to superdiffusive scaling, following Ref.~\cite{PhysRevLett.127.057201}. The only ingredient we will need that was not already motivated in elementary terms is the density of quasiparticles of size $s$ in an infinite-temperature state, $\rho_s \sim 1/s^3$. As we will see below, that scaling can also be understood from an elementary consistency argument; however, for the present let us assume this result. Then one can write the Kubo formula as a sum over strings, of the form
\begin{equation}\label{heiskubo}
\langle \hat{J}(t) \hat{J}(0) \rangle \propto \sum\nolimits_s \rho_s |v_s m_s|^2 \exp(-t/\tau_s).
\end{equation}
In this expression, $v_s \sim 1/s$ is the velocity of the $s$-string, $m_s \sim s$ is its bare magnetization, and $\tau_s$ is the time for which the string retains its initial magnetization (i.e., how long it takes to encounter a bigger string). 
Note that since the string is demagnetized after some finite time, its \emph{time-averaged} magnetization is zero, consistent with our remark that $m^{\mathrm{dr}}_s = 0$ for any finite $s$. 
It remains to estimate $\tau_s$. Strings of size $s$ demagnetize when they hit yet bigger strings; the spatial density of these is $1/s^2$. Since the $s$-string moves at velocity $1/s$, the \emph{time} it takes to hit a larger string is therefore $\tau_s \sim s^3$. Evaluating the Kubo formula with this data, we see that $\langle \hat{J}(t) \hat{J}(0) \rangle \sim t^{-2/3}$, giving the conductivity scaling $\sigma(\omega) \sim \omega^{-1/3}$ that corresponds to the dynamical exponent $z = 3/2$. 

\subsubsection{Quasiparticle density at infinite temperature} 

The argument above used $\rho_s \sim 1/s^3$ as an input. While this result can be directly computed from the thermodynamic Bethe ansatz~\cite{NMKI19}, its scaling follows from elementary constraints on the high-temperature static susceptibility at small finite magnetization $h$, namely $\langle \hat{S}^z_i \hat{S}^z_i \rangle = 1/4$~\cite{2020arXiv200908425I}. If we try to compute this as a sum over strings, it takes the form $\sum_s \rho_s (m_s^{\mathrm{dr}})^2$. 
%
In the presence of a field $h$, the density of Goldstone modes of size $s \gg 1/h$ is suppressed as $\exp(-hs)$. Since these solitons are rare, processes where they collide with even larger solitons are exponentially suppressed and can be neglected. Therefore their dressed magnetization $m^{\mathrm{dr}}_s = s$. By contrast, the populations of Goldstone modes of size $s \leq 1/h$ are not suppressed by the net magnetization, and their dressed magnetization should vanish at half filling. We have already seen that near half filling the dressed magnetization scales as $h$. By continuity with the large-$s$ limit, the dressed magnetization for $s \ll 1/h$ must scale as $m_s^{\mathrm{dr}} \sim h s^2$. 
Let us assume that $\rho_s \sim 1/s^\alpha$ for $s \ll 1/h$. Then the susceptibility scales as $h^2 s^{5-\alpha}|_{s \sim 1/h}$. To make this approach a constant, we require $\alpha = 3$. This result is consistent with TBA calculations.

Many calculations on the Heisenberg spin chain apply the strategy used above, which is to calculate physical quantities at very small nonzero $h$ and take the $h \to 0$ limit at the end. Sums over strings of size $s \leq 1/h$ are ubiquitous in these calculations. In the limit $h \to 0$, we can define a rescaled variable $\eta \equiv sh$ in terms of which these sums become Riemann sums, that can then be expressed as integrals over $\eta$~\cite{PhysRevLett.125.070601}. In this scaling limit, the TBA equations for the quantum spin chain and the classical Ishimori spin chain~\cite{Ishimori1982} coincide. The main difference between the two sets of equations is that in the classical problem $s$ is not quantized: thus, the classical and quantum models differ in their small quasiparticles but not in their large, long-wavelength fluctuations (as one would expect on general grounds). 

\subsection{Relation to KPZ scaling}

We now present an alternative argument for $z = 3/2$ scaling, which is slightly less direct but illuminates the relationship between superdiffusion in the Heisenberg model and Kardar-Parisi-Zhang (KPZ) scaling~\cite{gvw}. Even when the magnetization is zero on average, its instantaneous local value fluctuates with a variance given by the susceptibility, which is $O(1)$. Therefore we can think of the system, coarse-grained to any scale, as a patchwork of locally slightly magnetized regions. In each region, the magnetization is carried by the largest available string, which is of size $s \sim 1/h$ and has characteristic velocity $v_s \sim h$. If we are interested in magnetization transport over a distance $\ell$, then the characteristic value of $h \sim 1/\sqrt{\ell}$, so $v(\ell) \sim 1/\sqrt{\ell}$. Thus the timescale for spin transport over distance $\ell$ is $t(\ell) \sim \ell / (1/\sqrt{\ell}) \sim \ell^{3/2}$.

The observation that the local velocity associated with spin transport scales as $h$ is highly suggestive of the Burgers equation, $\partial_t m + \lambda \partial_x m^2 + D \partial^2_x m = \partial_x \xi$, where $\xi$ is white noise. (In our case the white noise would come from local thermal fluctuations of the smaller quasiparticles, which impart random scattering shifts~\cite{ghkv}.) The Burgers equation, in turn, is equivalent to the KPZ equation under a change of variables~\cite{kpz}. Numerical studies of the dynamical spin structure factor, in both quantum~\cite{PhysRevLett.122.210602} and classical~\cite{PhysRevE.100.042116, 1909.03799} systems, support the KPZ form $C(x,t) = t^{-2/3} f_{\mathrm{KPZ}}(x/t^{2/3})$, where $f_{\mathrm{KPZ}}$ is a universal scaling function computed in Ref.~\cite{Prahofer2004}. A detailed analytic argument supporting this conclusion was put forward in Ref.~\cite{vir2019}. However, this argument relies on assumptions that are not yet firmly established, and a complete analytic derivation of KPZ scaling in this model remains an open question.





\subsection{Away from integrability: Goldstone mode physics}

The fact that the large solitons that dominate high-temperature spin transport are Goldstone-mode wavepackets has important consequences for dynamics slightly away from integrability. Suppose one adds an integrability-breaking perturbation such as weak spatio-temporal fluctuations in the Heisenberg coupling. Since the noise acts \emph{locally}, it couples weakly to very large strings, since these locally look like vacuum rotations. Since the energy of an $s$-string scales as $1/s$, noise that couples to the energy can have at best a matrix element $\sim 1/s$, giving that string a decay rate $\sim 1/s^2$. (Another way to say this is that noise couples to Goldstone modes with a gradient.) Plugging in this decay rate $\tau_s \sim s^2$ into the Kubo formula~\eqref{heiskubo}, we find that the a.c. conductivity \emph{diverges} at low frequencies, as $\sigma(\omega) \sim |\log\omega|$. This prediction---and more generally the idea that there is a distinction between noise that preserves $SU(2)$ and noise that does not---are borne out by numerical studies~\cite{PhysRevLett.127.057201,2022PhRvB.105j0403M,roy2022robustness,mcroberts2022long}. 

This observation about the robustness of Goldstone modes against noise raises an intriguing question: while anomalous transport persists at lowest order in perturbation theory, is it present at arbitrary noise strength? The answer to this appears to be negative: both perturbation theory in the strong-noise limit~\cite{PhysRevLett.128.246603} and general hydrodynamic arguments~\cite{2020arXiv200713753G} suggest that transport should be simply diffusive when the system is noisy enough to be far from integrability. Since it seems unlikely that there is a transition as a function of noise strength, the most plausible way of reconciling the weak-noise and strong-noise results is to posit higher-order or nonperturbative effects that restore diffusion. What these effects (and the associated timescales) might be remains an open question. 

While the case of time-independent, Hamiltonian integrability-breaking perturbations (such as a next-nearest neighbor coupling) is much harder to study numerically, there has been interesting analytic progress on this problem~\cite{PhysRevB.105.104302}. The essential observation is that one can perform a unitary rotation $\hat U$ on $\hat H$ to extract $ \tilde{H} = \hat U(\lambda) \hat H \hat U^\dagger(\lambda)$ such that, to leading order in $\lambda$, $ \tilde H = \hat H + \lambda \hat V$ where $\hat V$ is the integrability-breaking perturbation. While such a $\hat U \equiv e^{i\lambda \hat S}$ can always be found, in general $\hat S$ has an exponentially large norm and the expansion is invalid~\cite{PhysRevX.10.041017}. However, for few-body Hamiltonian perturbations of the Heisenberg model, it turns out that many natural perturbations are generated by bounded, quasi-local operators $\hat S$~\cite{PhysRevB.105.104302, PhysRevX.5.041043}. Whenever this is the case to linear order in $\lambda$, the decay rate of conserved quantities is at best $\lambda^4$ (which is well out of reach of numerics) and possibly even slower~\footnote{I. L. Aleiner, private communication.}. In any case, large-scale numerical studies on classical models yield dynamics that is superdiffusive out to the longest accessible timescales~\cite{roy2022robustness, mcroberts2022long}. 

\subsection{Away from linear response}

Although theory is simplest in the linear-response limit, experiments involve finite-strength perturbations. Often, these perturbations are relatively strong, in order to enhance the signal-to-noise ratio.
Since the Heisenberg model is a sort of ``dynamical critical point'' between the ballistic and diffusive phases, one would generically expect its nonlinear response to be singular in the low-frequency limit. 
Since the dynamics has no intrinsic scale, the conductivity must be cut off either by $\omega$ or by the scale set by the perturbation strength, whichever is larger. In the latter case, we expect that the response to a perturbation is nonanalytic in the perturbation strength. 
There are many different protocols for probing nonlinear response; below, we review three specific settings that have been discussed in the literature. 
It will become apparent that even for these settings there are many open questions.

\subsubsection{Large domain walls}

The first instance of nonlinear transport we consider involves initializing two separate semi-infinite reservoirs at opposite chemical potentials $\pm \mu$ at time $t = 0$. In the previous section we discussed this setup for the easy-axis XXZ model and presented a closed-form expression for the full counting statistics of charge transport. At the Heisenberg point this approach breaks down, but one can nevertheless draw some conclusions about the \emph{mean} transport properties. 

Let us sketch the analysis of this setup using GHD. One can characterize each half-chain by its quasiparticle distribution. At finite $\mu$, we have that $\rho_s$ is exponentially suppressed for $s > 1/\mu$. Since the two half-chains have different quasiparticle vacua, when we combine the two systems we must also describe the spatial variation of the vacuum orientation---either by positing some dynamics for the vacuum orientation~\cite{vir2019} or by modeling the vacuum as a giant quasiparticle, as we did while discussing full counting statistics in the easy-axis phase. For now let us take the latter perspective. Then, as in the easy-axis phase, we get a diffusive contribution to transport whenever a quasiparticle is transmitted across the domain wall. A straightforward calculation shows that this is
\begin{equation}
D(\mu) \sim \sum_s \int d\lambda \rho_s(\lambda) (m_s^{\mathrm{dr}})^2 |v_s(\lambda)| \sim 1/\mu.
\end{equation}
Thus, for any finite density step, superdiffusion eventually gives way to diffusion, on a timescale that can be estimated (by equating $t^{2/3}$ and $\sqrt{t/\mu}$) to be $t^* \sim 1/\mu^3$. Numerical studies of classical spin chains~\cite{PhysRevLett.128.090604} support the scenario of an eventual crossover to diffusion at finite $\mu$, but the predicted scaling with $\mu$ has not yet been established numerically. 

Although the conclusion of the argument above (that superdiffusion is cut off on a timescale set by $\mu$) is plausible, the analysis is incomplete. To see why, let us go to the limit of a fully polarized domain wall, where the two half-chains are initially in opposite-spin vacua. The logic above would imply that this initial state remains frozen for all times. While this implication holds in the easy-axis phase, it fails at the isotropic point. An explicit solution~\cite{PhysRevB.99.140301} for this class of initial states is available using inverse-scattering methods, and shows that a domain wall smears out over a distance $x \sim \sqrt{t \log t}$. In terms of the quasiparticle picture, this can be understood by noting that quasiparticles in the Heisenberg model are smooth vacuum rotations rather than sharp domains; thus, our description of the initial condition (two identical quasiparticle distributions plus a giant quasiparticle on the left) is missing some of the quasiparticle content. 

At present, it remains an open question how one would incorporate these effects into GHD. A related open question is how to compute the full counting statistics of transferred charge. 
As discussed below, experiments~\cite{wei2022quantum} have measured the first three moments of the magnetization-transfer distribution for fairly large $\mu$. The skewness of this distribution is large and essentially nondecreasing with time; its numerical value is close to the prediction for the KPZ universality class. In addition, both experiments and numerical simulations show good agreement with superdiffusive scaling out to moderately long times $t \approx 50$, even for large $\mu$ where the GHD argument above would predict that the system becomes diffusive at short times. At present these observations remain unexplained.

\subsubsection{Response to a strong applied field}

Instead of setting up an initial condition with a finite bias, one could imagine applying a strong field (i.e., a magnetic field gradient $H = \mathcal{E} \sum_n n S^z_n$) to the system for a finite time interval $\Delta t$ and measuring the induced current. This calculation is straightforward to do away from half filling at field $h$ (though the full answer in the half filling limit is not known). We briefly summarize the findings and refer to Ref.~\cite{fava2021hydrodynamic} for details. Each quasiparticle receives an impulse of size $\mathcal{E} \Delta t m^{\mathrm{dr}}_s$; this modifies the quasiparticle distribution function by creating an imbalance between left and right movers, leading to a net current. The current can be written as a sum over strings $s \leq 1/h$ as follows (we have defined $\varphi \equiv \mathcal{E} \Delta t$):
\begin{equation}
J(h, \varphi) = h \sum_{s = 1}^{1/h} s^{-4} f(h \varphi s^3).
\end{equation}
Here, $f$ is a bounded periodic function of its argument, and for small argument $f(x) \approx x$. There are two regimes of response. First, when $\varphi \to 0$ at finite $h$, the argument of $f(x)$ remains small and the function can be expanded in a Taylor series in $\varphi$, giving a hierarchy of well-defined nonlinear response coefficients. While the linear-response coefficient (i.e., the term linear in $\varphi$) is regular as $h \to 0$, all higher-order coefficients are singular as $h \to 0$. To understand where these singularities come from, it helps to consider the limit $h \to 0$ at small but finite $\varphi$. Effectively, in this limit, the sum over strings is cut off at $h \varphi s^3 \approx 1$. As a consequence, $J(h, \varphi) \sim h^2 \varphi |\log (h\varphi)|$---the response becomes singular in the impulse in the superdiffusive limit. Physically, the mechanism for this effect is that large strings respond so strongly to the impulse that they effectively undergo Bloch oscillations, and thus do not contribute to response. The impulse $\varphi$ sets the boundary between strings that do and do not contribute to response. 

At present, nonlinear response has only been calculated using Euler-scale methods away from half filling. One expects an even more strongly nonanalytic response precisely at half filling, due to effects that are beyond the Euler-scale analysis. Performing this calculation would require a framework for computing nonlinear response beyond Euler scale, which remains an open problem. 

\subsubsection{Spin helices}\label{sechelix}

In ultracold atomic gases, it is often more convenient to prepare a pure state than a mixed one. Spin helices~\cite{bloch_heisenberg,Jepsen:2020aa}---in which the spin orientation varies in a spatially periodic manner, with wavelength $2\pi/k$---form a particularly natural class of initial states (see Sec.~\ref{sec:expt}). Spin helices are, in general, quite far from being locally thermal states, so the applicability of GHD to them can be questioned. However, one might expect that after some initial evolution, dephasing leads to local reduced density matrices that are well described by GGEs~\cite{Jacopoetal}. This intuition can be verified for free fermions. On the other hand, at present it is not entirely clear how to construct the appropriate GGEs for strongly interacting systems like the Heisenberg spin chain. 

One can define a length-time scaling form for the spiral initial state by measuring the wavenumber-dependence of the decay rate $\Gamma(k)$: a uniform state does not decay, so we expect $\Gamma \sim k^{z}$. Since the spiral is very far from equilibrium, there is no guarantee that this dynamical exponent will coincide with the linear-response one. Evidence from experiments, short-time numerical simulations, and the short-time expansion suggests that in the easy-plane regime, helices relax ballistically ($z = 1$). The situation in the easy-axis phase is much less clear, but in the large-$\Delta$ limit the spiral rapidly dephases under the Ising interaction, so on theoretical grounds we expect $z = 2$. Interestingly, at the isotropic point helices seem to relax with a diffusive dynamical exponent ($z = 2$), not the KPZ exponent. 
This case is delicate since (owing to the $SU(2)$ symmetry) the energy density of the initial state is itself proportional to $k^2$; thus, by changing $k$, one is tuning both the ``temperature'' and the length-scale. 
Another subtlety is that at long wavelengths the Heisenberg model can be regarded as coarse-graining into the Landau-Lifshitz model, for which the spiral is an exact eigenstate.
It remains to be seen whether these features can be incorporated into GHD.


\section{Easy-plane XXZ} \label{SecEasyPlane}

Finally, we discuss the easy-plane regime $\Delta <1$, where the XXZ spin chain is gapless at low energy, and is described by a Luttinger liquid. We will argue that spin transport is ballistic in this regime, albeit with a highly nontrivial ``fractal'', discontinuous Drude weight as a function of the anisotropy. This remarkable behavior of the Drude weight has interesting physical consequences for the low-frequency behavior of the conductivity, as the spectral weight must somehow rearrange itself as the anisotropy is changed. We also briefly discuss weak-integrability breaking in that phase. 

\begin{figure}
    \centering
   \includegraphics[width = 0.55\columnwidth]{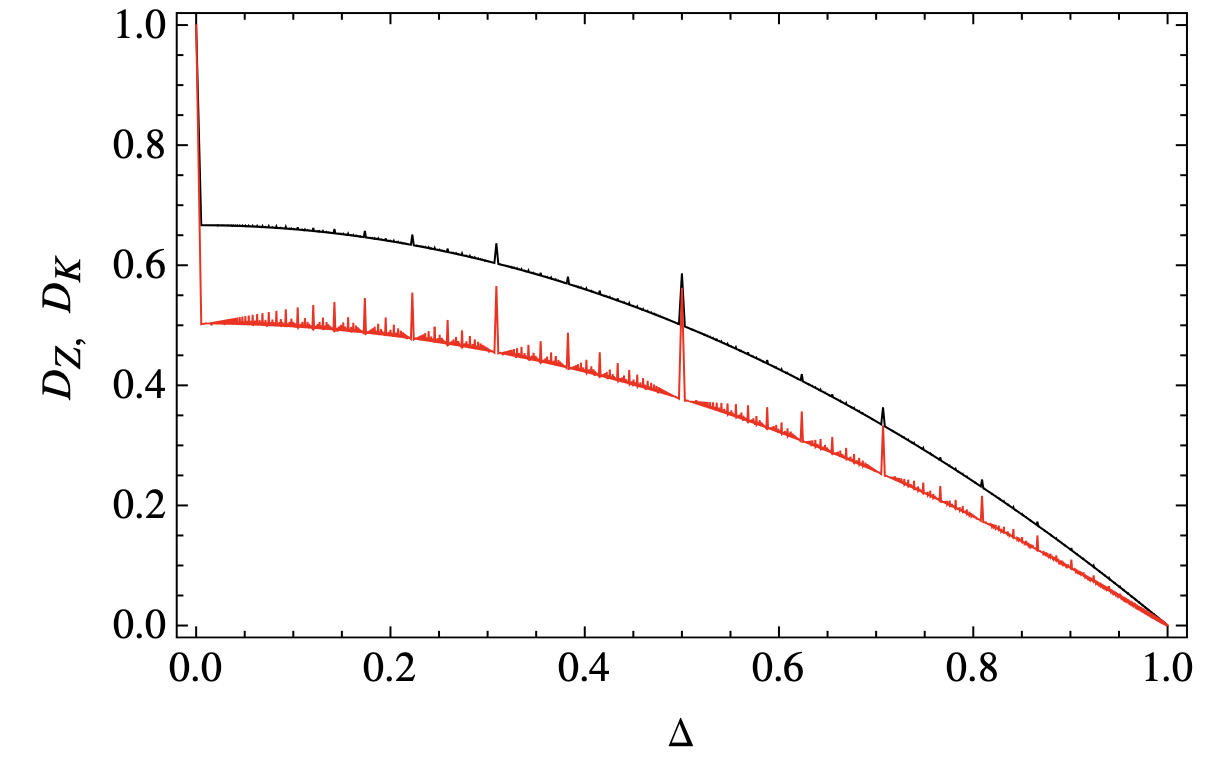}
    \caption{{\bf Fractal Drude weight.} Lower-bound on the Drude weight (red, from~\cite{PhysRevLett.106.217206}), and optimized bound in black (from~\cite{PhysRevLett.111.057203}), coinciding with GHD predictions. Reproduced from Ref.~\cite{PhysRevLett.111.057203}.   }
    \label{figDrudeWeight}
\end{figure}

\subsection{Quasiparticle content and fractal Drude weight}

The quasiparticle spectrum in the regime $ \Delta <1$ is much less intuitive than the other cases discussed above. In particular, the number of quasiparticle species depends on the value of the anisotropy $\Delta$ in a non-trivial, ``fractal'' fashion. More precisely, we write the anisotropy as $\Delta \equiv \cos \pi \lambda $. The quasiparticle (Bethe ``strings'') spectrum that follows from the Bethe ansatz solution is determined by the continued fraction expansion of $1/\lambda =\nu_1+\frac{1}{\nu_2+\frac{1}{\nu_3+...}}$. The total number of strings is given by  $n=\sum \nu_i$. For an irrational value of $1/\lambda$, the number of strings is infinite, while approximating it by a rational number leads to a finite number of strings~\cite{Takahashi}. 

This peculiar dependence of the quasiparticle content on the anisotropy has direct consequences on physical observables. The most striking example is the spin Drude weight. The quasiparticles have linear dispersion ($z=1$) and a  velocity that remains finite even for the largest ones, contrary to the $\Delta \geq 1$ case where infinitely large strings had vanishing velocity. As those large quasiparticles carry some dressed magnetization, spin transport is ballistic with a delta-function contribution in the conductivity characterized by a Drude weight $D_\lambda$. Remarkably, this Drude weight appears to be discontinuous and fractal as a function of $\lambda$. We will focus on the infinite temperature limit for concreteness. When $\lambda = p/q$ is rational, several distinct methods~\cite{PhysRevLett.82.1764, karraschdrude, Prosen20141177,PhysRevB.96.081118,PhysRevLett.111.057203,PhysRevLett.119.020602, urichuk2019spin} predict that
\begin{equation}
\label{drudeweight}
D_\lambda = \frac{1}{12} (1 - \Delta^2) f\left(\frac{\pi}{q}\right), \quad f(x) = \frac{3}{2} \left[ \frac{1 - \frac{\sin(2 x)}{2x}} { \sin^2 x } \right].
\end{equation}
Eq.~\eqref{drudeweight} is a rigorous lower bound on $\mathcal{D}$, which GHD~\cite{PhysRevB.97.081111, PhysRevLett.119.020602, BBH, PhysRevB.96.081118} predicts is saturated. 
Very surprisingly, Eq.~(\ref{drudeweight}) implies that the Drude weight jumps by ${\cal O}(1)$ as $\Delta$ changes infinitesimally (Fig~\ref{figDrudeWeight}): $\lim_{x \to 0} f(x) = 1$ for any irrational number $\lambda$, but is higher by an ${\cal O}(1)$ amount at an arbitrarily close small-denominator rational. This is a direct consequence of the fractal quasiparticle content as a function of $\lambda$. 

\begin{figure}
    \centering
   \includegraphics[width = 0.75\columnwidth]{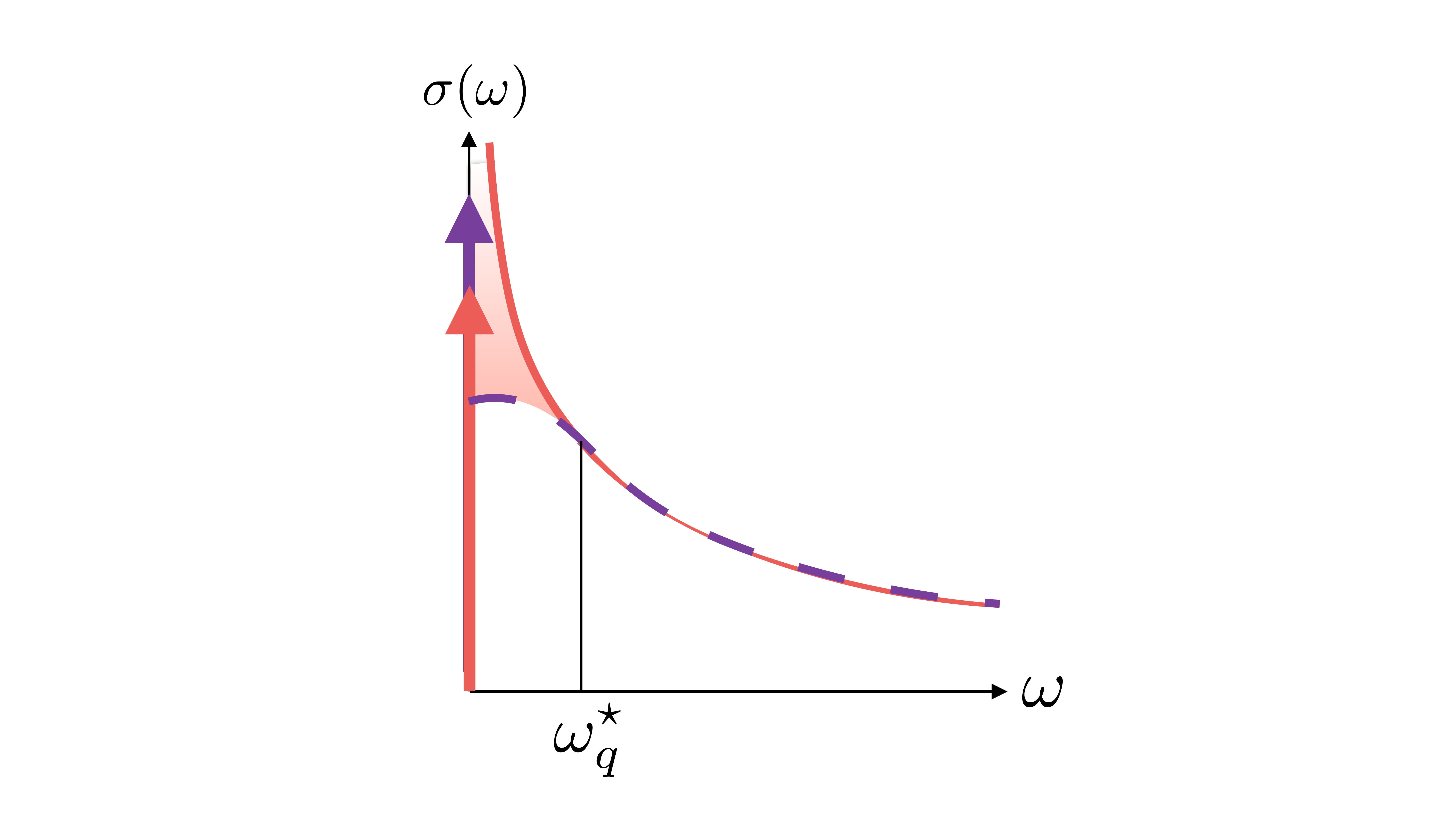}
    \caption{{\bf Anomalous corrections to ballistic transport for $\Delta<1$.} Conservation of spectral weight in the low-energy conductivity: the extra Drude weight at the rational point (purple) must precisely match the missing part of the regular spectral weight~\cite{2019arXiv190905263A}. As a result, the conductivity diverges for irrational values of $\lambda$, see eq.~(\ref{eqsqrtdiv}).      }
    \label{figSketchGapless}
\end{figure}

\subsection{Corrections to ballistic transport}

These discontinuous jumps in the zero-frequency spectral weight suggest that the finite-frequency behavior must also be nontrivial~\cite{2019arXiv190905263A}. By locality, correlators at finite $t$ must be continuous functions of the anisotropy $\Delta$ (or $\lambda$), since they only probe system sizes of order $\sim t$. This means that if we take two anisotropies that differ by a small value $\epsilon$, we must go to long times or low frequencies $< \epsilon$ to distinguish them. This simple observation strongly constraints the low-energy scaling of the conductivity. Let us consider an irrational value $\lambda_\infty$, that we approximate by a sequence of rationals $\{ \lambda_q = p/q \}$ with increasing denominators $q$. Above a crossover frequency scale $\omega_q^\star$, the conductivity for the rational approximant and the irrational value are identical. The rearrangement of the spectral weight due to the change in the Drude weight~(\ref{drudeweight}) $\delta D_q \equiv D_{\lambda_q} - D_{\lambda_\infty}  \sim 1/q^2$ must occur at low frequencies $\omega < \omega^\star_q$. GHD predicts that for rational $\lambda$ (finite $q$), the regular part of the conductivity (excluding the Drude weight) approaches a finite~d.c.~value at low energy: $\sigma^{\rm d.c.}_{\lambda_q} \sim q^2$. This blows up for irrational values of $\lambda$, and we assume an algebraic divergence with frequency $\sigma_{\lambda_\infty}(\omega) \sim \omega^{-\alpha}$. Conservation of spectral weight then implies that $\int_{0}^{\omega^\star_q} d\omega [\sigma_{\lambda_\infty}(\omega) - \sigma^{\rm d.c.}_{\lambda_q}] \simeq \delta {\cal D}_q \sim 1/q^2$: the extra Drude weight at the rational point must precisely match the missing part of the regular spectral weight~\cite{2019arXiv190905263A} (Fig.~\ref{figSketchGapless}). This means that $\int_{0}^{\omega^\star_q} d\omega \sigma_{\lambda_\infty}(\omega) \sim (\omega^\star_q)^{1-\alpha}$ and $\int_{0}^{\omega^\star_q} d\omega \sigma^{\rm d.c.}_{\lambda_q} \sim \omega^\star_q q^2 $ must both be of order $q^{-2}$ to compensate each other. This yields $\omega^\star_q \sim q^{-4}$, and $\alpha=1/2$, so that~\cite{2019arXiv190905263A} 
\begin{equation} \label{eqsqrtdiv}
\sigma_{\lambda_\infty}(\omega) \sim \frac{1}{\sqrt{\omega}}. 
\end{equation}
We see that conservation of spectral weight combined with the fractal nature of the Drude weight implies that the corrections to ballistic transport are {\em superdiffusive}, with dynamical exponent $z=4/3$ -- recall that in general, $\sigma(\omega) \sim \omega^{1-2/z}$. 

 The crossover scale $\omega^*_q \sim q^{-4} \sim |\lambda_\infty - \lambda_q|^2$ resembles a Golden Rule rate, suggesting the following natural interpretation: if one starts in an eigenstate of $\hat H_{\lambda_\infty}$ and turns on the perturbation $\delta \hat  H \equiv \hat H_{\lambda_q} - \hat H_{\lambda_\infty}$, the quasiparticle structure changes considerably. The quasiparticles at $\hat H_{\lambda_\infty}$ are no longer stable, and it is natural to suppose that their decay rates scale as $|\lambda_q - \lambda_\infty|^2$, yielding $\omega^*_q \sim q^{-4}$.


\subsection{Away from integrability}

Moving away from integrability is somewhat more conventional in this phase. Quasiparticles acquire a mean-free path and time, over which they decay and/or backscatter. The Drude weights is broadened into a generalized Lorentzian whose width is given by the inverse mean free time~\cite{friedman2019diffusive,2020arXiv200411030D}, as in the familiar case of weakly interacting fermions. However, this can lead to non-analytic behavior of the resulting diffusion constant as a function of the integrability breaking perturbation~\cite{friedman2019diffusive,PhysRevLett.125.180605}. Starting from $\sigma(\omega) = \pi D \delta(\omega) + C \omega^{-1/2}$ in the integrable case, we expect a finite d.c.~conductivity  $\sigma = \pi D \tau + C \tau^{1/2}$ away from integrability.

\section{Numerics and experiments} \label{SecExp}

We now turn to numerical and experimental tests of the GHD-based theory of anomalous transport outlined above. 
In principle, GHD is a theory of late-time behavior in an infinite system, while both numerics and experiment are limited to finite times and/or finite system sizes (which restrict the GHD regime to finite times, see Sec.~\ref{limits}). 
However, in practice the GHD regime in XXZ spin chains sets in at times of order unity. Therefore, exact numerical simulations in particular have been a valuable consistency check on GHD predictions, and in particular on the validity of extensions of GHD to treat anomalous transport in spin chains.
In addition, exact results from GHD can be used a stringent test of the validity of approximate classical simulations, or of quantum simulations using noisy quantum computers. 

\subsection{Numerical approaches}

\subsubsection{Tensor-network operator evolution} 

The most direct approach to linear-response transport is to explicitly evaluate the Kubo formula. This involves computing time-evolved operators in the Heisenberg representation. At short times such operators can be represented as matrix-product operators (MPOs), and the time-evolution of the MPOs can be computed using time-evolving block decimation (TEBD)~\cite{PhysRevLett.91.147902}. By the Lieb-Robinson theorem, the time-evolved operator lives inside the light-cone, so these calculations effectively take place in the thermodynamic limit if the system size is larger than the light cone at the latest accessible time. In practice TEBD is exact only up to modest times ($t \leq 25$) before the MPO required to accurately represent the time evolved operator becomes too complex to work with (this quantity is called the ``bond dimension'' of the MPO). At present, it is unclear whether the bond dimension asymptotically grows polynomially or exponentially with system size. For existing computational resources, TEBD is limited to similar time windows as the other available methods. 


\subsubsection{Density matrix evolution from weak steps}\label{ljubot}

Instead of working in the Heisenberg picture, one could instead have started in the Schr\"odinger picture, with an initial density matrix of the form $\hat \rho = \prod_i (1 + \mu\mathrm{sgn}(i) \hat\sigma^z_i)$. This represents an initial state with a step profile of the density. One lets this initial state evolve for time $t$ and then measures $Z(x,t) \equiv \mathrm{Tr}(\hat \sigma^z_x \hat \rho(t))$. In the limit $\mu \to 0$ a straightforward calculation~\cite{PhysRevLett.122.210602, lzp} shows that the structure factor $C(x,t) = \partial_x Z(x,t)$. Like the previous approach, the step profile approach has the appealing feature that the density matrix does not evolve outside a light cone emanating from the domain wall (since away from the domain wall the initial state is a stationary state of the dynamics). Once again, one represents the initial density matrix as an MPO and evolves it using TEBD. A major difference between this and the previous approach is that the step initial profile need not be in the linear-response regime: this is an advantage if one wants to go beyond linear response, but can also have drawbacks if one wants to establish the linear-response answer, since this requires one to compute $Z(t)$ as a function of $\mu$ and extrapolate to the $\mu \to 0$ limit. 

For reasons that are not fully understood, fixing a bond dimension, truncating the time-evolved density matrix to an MPO of that bond dimension, and evaluating $Z(t)$ gives physically sensible answers at surprisingly late times~\cite{lzp}. Whether the truncation causes quantitative errors remains unclear.

\subsubsection{Boundary Lindblad}

In a typical solid-state conductivity experiment, one does not directly time evolve the state, but instead connects it to leads at different chemical potentials and measures the current passing through it in its steady state. This steady-state setup can be directly simulated as follows~\cite{PhysRevLett.106.220601}. One considers the system coupled to Markovian leads at both boundaries. At the left boundary, particles are injected at some rate $\mu_+^L$ and absorbed at some rate $\mu^-_L$, and similarly at the right boundary. If more particles are injected at the left and absorbed at the right, then in the steady state there is (i) a net current flowing through the bulk of the chain, and (ii) a density gradient across the chain. The dependence of the steady-state current on the length of the chain allows one to extract the transport exponent, as follows. Define $V$ as the chemical potential difference across a system of size $L$. The current is related to this by $j = (V/L) \sigma(L)$, where $\sigma(L)$ is the conductivity for a system of size $L$. For this finite system, we can use the Einstein relation $\sigma(L) \sim D(L)$. Assuming a length-time scaling $t \sim L^z$, and using the fact that $L \sim \sqrt{D(L) t}$, one can derive the relation $j \sim L^{1-z}$. 
Under the (numerically borne-out) assumption that the steady-state density matrix can be represented as a low-bond-dimension MPO, one can find the steady state for large systems by variationally minimizing the norm $\Vert \mathcal{L}(\hat\rho)\Vert$ across MPOs. 

The boundary Lindblad method is a powerful approach that has led to many new results on the dynamics of integrable systems, including the first identification of $z = 3/2$ scaling~\cite{PhysRevLett.106.220601}. However, interpreting these results in terms of GHD requires caution: GHD addresses the regime of times much less than system size, while the boundary Lindblad technique finds the slowest mode in a finite-size system. In general these limits do not commute (Sec.~\ref{limits}). An example where one sees a discrepancy is at large $\Delta$: direct time evolution is consistent with the GHD result that the diffusion constant saturates to a nonzero value as $\Delta \to \infty$, whereas the boundary Lindblad approach shows a vanishing diffusion constant~\cite{PhysRevLett.106.220601}. A possible (but still speculative) explanation is that this is a consequence of breaking integrability at the boundary and going to very late times at finite system size. 

\subsubsection{Time evolution of pure states} 

A more direct method, which avoids truncation at the expense of introducing finite-size effects, is to replace the infinite-temperature state with a random vector (by canonical typicality) and time-evolve using sparse matrix-vector multiplication. Typicality methods were used to compute the optical conductivity of charge and energy~\cite{PhysRevLett.107.250602, hsp, PhysRevLett.116.017202} and subsequently to explore full counting statistics~\cite{2022arXiv220309526G}. The gain from working with pure states is that this approach avoids matrix multiplication, and instead only requires the sequential application of sparse matrices to the random initial state. Typicality approaches are ultimately limited by system size, since it is infeasible to store even a pure state on more than $L \approx 32$ qubits (though slightly bigger sizes are achievable using TPUs~\cite{PRXQuantum.3.020331}). 

For weak domain walls and linear-response calculations, the direct approach performs slightly worse than the MPO approaches discussed above. Its relative advantage is in dealing with general far-from-equilibrium pure states. For such states, TEBD can fail at much earlier times than direct evolution: state-of-the-art bond dimensions are limited to $\sim 2000$, which is two orders of magnitude lower than the bond dimension needed to describe a maximally entangled state on $36$ qubits.







\subsubsection{Approximate methods}

The past five years have seen the development of multiple numerical methods based on physically motivated truncations of MPS's and MPO's. Examples of these methods include the time-dependent variational principle (TDVP)~\cite{2017arXiv170208894L}, density-matrix truncation (DMT)~\cite{PhysRevB.97.035127}, and dissipation-assisted operator evolution (DAOE)~\cite{2020arXiv200405177R}. These methods aim to truncate the growth of entanglement in ways that do not affect the dynamics of simple observables, such as spin or energy. In practice, all of these truncation schemes break integrability, so it is unclear \emph{a priori} that they should be able to faithfully describe transport in integrable systems. Nevertheless, promising results on the Heisenberg spin chain have been achieved with DMT as well as DAOE~\cite{2020arXiv200405177R, ye2022universal}. One can regard our exact results on integrable systems as providing a benchmark to test the performance of these methods against.




\subsubsection{Away from integrability: noisy evolution}

Unitary evolution of operators in the Heisenberg picture is limited to early times for both integrable and nonintegrable spin chains. For nonintegrable systems, operator entanglement grows somewhat faster, and the accessible simulation times are correspondingly shorter ($t \leq 20$). This makes the numerical study of nearly integrable unitary dynamics challenging, since one generically expects integrability-breaking effects to become visible on timescales that scale as the \emph{square} of the integrability-breaking parameter. For this estimated Golden Rule decay rate to be visible in the numerically accessible window, one has to break integrability rather strongly, to a point where perturbation theory is potentially no longer valid. 

A more feasible way to study these effects is to sacrifice the conservation of energy, and instead consider perturbations that act randomly in time, i.e., as noise. In this case, one can represent the dynamics averaged over the noise as a quantum channel or Lindblad master equation~\cite{PhysRevLett.127.057201}. Operator entanglement growth under quantum channels is slow, by the following logic. Instead of thinking of the channel as a noise-average, one can equivalently think of it as the dynamics of an open system. Under open-system dynamics, operators that begin entirely inside the system spread into the environment. Larger operators have more sites where they can leak into the environment. Tracing over the environment annihilates these parts of the time-evolved operator, which therefore loses operator norm over time but remains weakly entangled at all times. Therefore, it is practical to simulate noisy evolution for much longer times ($t \geq 100$) than strictly unitary evolution. 

One might worry that noise is too crude a way to break integrability; however, as we discussed above, many of the interesting consequences of breaking integrability occur regardless of whether the perturbation is Hamiltonian.



\subsection{Experimental studies}\label{sec:expt}

\subsubsection{Magnetic materials probed by neutron scattering}

As we noted in the introduction, the study of high-temperature transport was revived by experiments in ultracold atomic gases and other forms of synthetic matter. Even in solid-state systems, the characteristic energy scales associated with \emph{magnetic} excitations can be quite low, however, allowing one to study high-temperature spin dynamics at absolute temperature scales that are quite low (i.e., substantially below room temperature). Many of the initial experiments addressing this question were concerned with establishing the presence of ballistic energy transport (which suggests integrability); accordingly, they focused on the large (apparently) magnetic contribution to the heat conductivity~\cite{bertini2020finite}.

While measuring d.c.~spin transport can be challenging, one can extract the dynamical spin structure factor $S(q,\omega)$ from neutron scattering. The first experimental evidence for superdiffusion in Heisenberg spin chains came from a neutron scattering experiment on the quasi-one-dimensional magnet KCuF$_3$~\cite{Scheie2021}. In this experiment, $S(q,\omega)$ was studied at fixed, very low $\omega$ as a function of $q$. Over a wide range of temperatures (from $70$ K to $200$ K), it was found that $S(q, \omega) \sim q^{-z}$ with values of $z$ ranging from $1.3$ to $1.6$, depending on the temperature and the fitting window. While these results do not conclusively show $z = 3/2$, they are at least consistent with it, and inconsistent with either ballistic or diffusive scaling. 

While KCuF$_3$ realizes the isotropic Heisenberg model, a variety of other quasi-one-dimensional magnets exist that realize anisotropic XXZ spin chains~\cite{PhysRevResearch.4.013111}. Exploring high-temperature spin transport in the anisotropic regime is an interesting task for future work.

\subsubsection{Nuclear magnetic resonance}

The dynamics of XXZ spin chains can also be probed using nuclear magnetic resonance (NMR) experiments. There are two conceptually distinct contexts for using NMR; we will discuss these separately. 

First, NMR can be used as a probe of \emph{electron} spin dynamics. In an NMR experiment, nuclei are placed in a static polarizing magnetic field and then excited using microwaves of frequency $\omega_0$. The microwave frequency selects which nuclei (if any) are excited. Assuming that nuclear spins decay by coupling to magnetic fluctuations, the rate at which nuclear spins depolarize (i.e., the $T_1$ time) is related as follows to the local autocorrelation function of the nearby electron spins~\cite{PhysRevB.100.094411, PhysRevLett.127.107201}:
\begin{equation}
T_1^{-1} \propto \mathrm{Re} \int_0^{\omega_0^{-1}} dt \, \langle \mathbf{\hat S}_0(t) \mathbf{\cdot \hat S}_0(0) \rangle.
\end{equation}
Thus the dependence of the $T_1$ time on $\omega_0$ is a diagnostic of the transport exponent governing the local autocorrelation function. In particular, if the low-frequency conductivity diverges with some power law $\alpha$, then we expect $1/T_1 \sim \omega_0^{-\alpha}$. A divergence of this kind was in fact seen decades ago in NMR studies of the Heisenberg magnet Sr$_2$CuO$_3$~\cite{PhysRevLett.87.247202}, but NMR response in the high-temperature regime merits further exploration in light of the results on anomalous transport reviewed here. 


In addition to using NMR techniques as a probe of electronic spins, one can also use these techniques to engineer and study systems of interacting \emph{nuclear} spins, as in recent work on fluorapatite~\cite{wei2019emergent}. The active degrees of freedom in these experiments are spins-$1/2$ of fluorine nuclei. Because of the anisotropy of the material, interchain couplings are suppressed by a factor of $1/40$ relative to intrachain nuclear couplings, so the system is effectively one-dimensional on the experimental timescales of $\leq 50$ (in units of $J$). NMR experiments naturally take place at high temperatures relative to the nuclear interaction scales ($10$ kHz $\sim 10^{-7}$ K).
However, measuring transport in this setting has been challenging, because of the lack of local addressability. 

One traditional procedure for studying spin diffusion~\cite{PhysRevLett.80.1324} has been to create an initial state with an imprinted periodically modulated magnetization profile (by applying a linear magnetization gradient, discussed further below in Sec.~\ref{wolfgang}), and track how the contrast of this imprinted spin spiral decays. 
A different (and in some ways more versatile) approach to studying spin transport was very recently explored in Ref.~\cite{paipeng}. The idea is as follows. The ``intrinsic'' Hamiltonian for fluorine nuclear spins in fluorapatite takes the form
\begin{eqnarray}
\hat H & = & \hat H_{\mathrm{FF}} + \hat H_{\mathrm{dis}}. \nonumber \\
\hat H_{\mathrm{FF}} & = & \sum_{j < k} \frac{1}{|j - k|^3} (2 \hat{S}^z_j \hat{S}^z_k - \hat{S}^x_j \hat{S}^x_k - \hat{S}^y_j \hat{S}^y_k). \nonumber \\
\hat H_{\mathrm{dis}} & = & \sum_j h_j \hat{S}^z_j.
\end{eqnarray}
This Hamiltonian consists of two parts: the interactions among the fluorine nuclear spins, and the interactions of each spin with its random local environment (consisting mainly of phosphorus spins), which acts as a static random field. We will think of $\hat H_{\mathrm{FF}}$ as effectively local---justified given the short experimental timescales. Dynamical decoupling techniques exist that can temporarily remove either $\hat H_{\mathrm{FF}}$ or $\hat H_{\mathrm{dis}}$ to any desired precision~\cite{RevModPhys.76.1037}. In addition, established Floquet engineering techniques let one construct arbitrary XXZ Hamiltonians from $\hat H_{\mathrm{FF}}$ by applying particular pulse sequences~\cite{PhysRev.175.453}. 

The scheme of Ref.~\cite{paipeng} uses these features as follows (we have simplified the treatment a little to focus on the main ideas). First, one applies a strong uniform field along the $z$ direction, creating the state $\hat\rho \propto (\mathbb{I}_n + \epsilon \hat S^z_n)^{\otimes n}$. Second, one evolves under $\hat H_{\mathrm{dis}}$ (while decoupling $\hat H_{\mathrm{FF}}$) in order to reach a state where each spin has a random polarization along the $z$ axis: $\hat \rho \propto (\mathbb{I}_n + \alpha_n \hat S^z_n)^{\otimes n}$, where $\epsilon_n$ is an effectively random number. Third, one time evolves under the desired Floquet-engineered XXZ Hamiltonian (while decoupling $\hat H_{\mathrm{dis}}$). Fourth, one again decouples $\hat H_{\mathrm{FF}}$ and implements the second step in reverse, and finally measures the total polarization along the $z$ axis. This entire process extracts the observable
\begin{equation}
C(t) \equiv \mathrm{Tr}\left[ \sum\nolimits_{ij} \alpha_i \alpha_j \langle \hat S^z_i(t) \hat S^z_j \rangle \right],
\end{equation}
where $\alpha_i$ is a random, statistically translation-invariant, variable with zero average (sampled over many runs of the experiment). NMR experiments can therefore extract the sample-averaged local autocorrelator $\overline{C(t)} \propto \langle \hat S^z_i(t) \hat S^z_i(0) \rangle$. The results~\cite{paipeng} provide convincing evidence that interacting integrable systems can have ballistic energy transport coexisting with diffusive spin transport, unlike either nonintegrable systems or free systems.

From the perspective of exploring high-temperature dynamics, one valuable aspect of both NMR techniques discussed above is that they give access to the spatially local autocorrelation function. For $\Delta > 1$, as we discussed above (Sec.~\ref{secgvw}), this correlation function decays with an anomalous, continuously varying power law even in systems where transport is ballistic~\cite{gvw}. 


\subsubsection{Ultracold atomic gases}\label{wolfgang}

The revival of interest in integrability owes much to experiments on one-dimensional ultracold gases in the continuum~\cite{kinoshita}. These experiments demonstrated, in a striking way, that integrability has physically relevant consequences---it causes far-from-equilibrium initial states to remain out of equilibrium for extremely long times. 

Although these continuum experiments have made remarkable progress in testing GHD~\cite{PhysRevLett.122.090601, 2020arXiv200906651M, PhysRevLett.126.090602, Bouchoule_2022}, they generally realize the Lieb-Liniger model, in which transport is ballistic rather than anomalous. Integrable \emph{spin chains} are most naturally realized on the lattice. One first realizes a Fermi- or Bose-Hubbard model on a one-dimensional lattice by confining an ultracold atomic gas in a strongly anisotropic three-dimensional optical lattice, such that the lattice along two directions is so deep as to be essentially infinite. Along the third direction, the lattice is deep enough to be in the single-band tight-binding regime, but otherwise tunneling is allowed. In the experiments of Refs.~\cite{Jepsen:2020aa, wei2022quantum} the Hamiltonian governing the system is the Bose-Hubbard Hamiltonian
\begin{equation}
\hat H = - \sum_{i,\sigma} (\hat b^\dagger_{i,\sigma} \hat b_{i+1,\sigma} + \mathrm{h.c.}) + U_{\sigma\tau} \hat n_{i\sigma} \hat n_{i\tau}.
\end{equation}
Here, Greek indices denote the internal state (``spin'') of the atom. One could write a similar Hamiltonian for fermions, except that by the exclusion principle $U$ only affects pairs of atoms with opposite spin. The Fermi-Hubbard model is itself integrable in one dimension. The Bose-Hubbard model is not integrable, but in the limit $U_{\sigma\tau} \gg 1$, the system enters a Mott insulating phase in which charge fluctuations are suppressed and the only remaining dynamics is the spin dynamics due to superexchange. This is governed by the Hamiltonian~\cite{Jepsen:2020aa}
\begin{equation}
\hat H_{\mathrm{XXZ}} = \sum_i - \frac{4}{U_{\uparrow\downarrow}} (\hat S^x_i \hat S^x_{i+1} + \hat S^y_i \hat S^y_{i+1}) + \left( \frac{4}{U_{\uparrow\downarrow}}- \frac{4}{U_{\uparrow\uparrow}} - \frac{4}{U_{\downarrow\downarrow}} \right) \hat S^z_i \hat S^z_{i+1},
\end{equation}
up to an overall field that does not affect the dynamics. Absent fine-tuning, all the $U_{\sigma\tau}$ tend to be similar in magnitude, so $\Delta \approx 1$. However, by tuning the strength of one of the on-site interactions using a Feshbach resonance, one can realize strongly anisotropic XXZ models~\cite{Jepsen:2020aa}. Whether such Feshbach resonances are available depends on the atomic species: rubidium has few usable resonances in the experimentally relevant parameter range, whereas lithium and cesium have many. 

In general, there will also be defects in the filling of the lattice, and these will give rise to residual charge dynamics that breaks integrability. On the relatively short timescales that have been experimentally studied, these effects do not seem to break integrability, but do seem to renormalize the dynamical timescales~\cite{wei2022quantum}. 

Two main strategies have been adopted to initialize the effective XXZ spin chain and study its dynamics. The conceptually simpler approach requires a quantum gas microscope with single-site resolution. Here, one can ``dial in'' a variety of initial states. In the experiments of Ref.~\cite{wei2022quantum}, the initial state was a random product state with net magnetization density $\pm \eta/2$ on the left (right) half of the chain. This setup closely parallels the numerical setup discussed above in Sec.~\ref{ljubot}. After letting the system evolve, one can take a simultaneous snapshot of all the spins in the system; both average transport and fluctuations can be read off from the statistics of these snapshots. These experiments used rubidium atoms and were therefore restricted to the isotropic limit. The average transport exponent was consistent with $z = 3/2$; the fluctuations were anomalous, as discussed in Sec.~\ref{SecXXX}.

In the experiments of Ref.~\cite{Jepsen:2020aa}, which did not use a quantum gas microscope, the system was instead initialized in a spin spiral of period $2\pi/q$, rotating in the XZ plane, using a magnetic field gradient. %
After some evolution time the spiral was unwound and its residual contrast was read out.
This method is restricted to a specific class of initial states; however, in principle, one can read out the full correlation function $C(x,t) = \langle \hat S^z_x(t) \hat S^z_0(t)\rangle$ as a function of time after the quench (i.e., one can wind the spiral up at pitch $q$ and unwind it at pitch $q'$, thus getting the full instantaneous structure factor of the time evolved state). 

Since the initial state is far from equilibrium, its dynamical properties are very different from linear response transport (Sec.~\ref{sechelix}), and many questions still remain. 

\subsubsection{Superconducting circuits}

In the past decade, arrays of superconducting qubits have emerged as a platform for the digital simulation of many-body systems~\cite{krantz2019quantum}. These arrays can implement a variety of one- and two-qubit gates, including those needed for simulating the integrable Trotterization of the XXZ spin chain. The geometry is tunable, allowing for controlled studies of dimensional crossover in quasi-1D systems. Some basic features of the Trotterized XXZ spin chain, including the presence of stable string states in the spectrum~\cite{morvan2022formation} and of nearly conserved quantities~\cite{maruyoshi2022conserved}, have recently been established in these systems. These works exploited some of the distinctive capabilities of superconducting circuits: in particular, the ability to measure arbitrary Pauli strings. However, transport from initial finite-entropy states has not yet been explored in this setting. 

At present, the lifetimes of superconducting circuit experiments are somewhat lower than those of experiments with ultracold atoms. Experiments in superconducting circuits are limited by finite single-qubit decay ($T_1$) and decoherence ($T_2$) times, as well as errors due to finite gate fidelity. All of these errors except for the $T_1$ process are integrability-breaking perturbations that preserve the $U(1)$ symmetry; therefore, one can study the properties of transport in nearly integrable systems without mitigating or correcting these errors. Qubit decay breaks the $U(1)$ symmetry; however, since all decay events increase the number of $\downarrow$ spins relative to the initial state, one can remove the effects of decay at the expense of post-selecting samples with the right number of $\uparrow$ spins. Asymptotically, however, samples with no decay events become exponentially rare in both system size and evolution time, so that the need to post-select limits the scalability of such experiments.
In state-of-the-art work, the stability of strings in the $\Delta > 1$ regime has been demonstrated for timescales that are on the order of $\approx 30$ layers of gates, on a system of $24$ qubits. A subsequent experiment~\cite{maruyoshi2022conserved} demonstrated that one could use this platform to perform direct measurements of conserved charges and how they evolve under noisy approximately integrable dynamics. Superconducting circuits are particularly well suited for such experiments since they allow for single-site measurements in arbitrary bases: therefore, one can measure arbitrary Pauli strings, as opposed to just strings in the computational basis. 

\section{Conclusion} \label{SecConclusion}

Integrable systems at finite temperature have unusual dynamical properties because they have stable ballistic quasiparticles whose properties are nontrivially renormalized by the interactions between them. In particular, interactions can strip quasiparticles of their charge, leading to sub-ballistic transport of the types we have discussed here. The picture of transport through thermal fluctuations that we have presented here was fleshed out in the past four years, and continues to see rapid development; as it stands, it is a relatively complete theory of the low-frequency limit of the $q \to 0$ linear-response conductivity. Beyond linear response, there are more questions than answers, but progress in extending the theory to this regime has been rapid. 

An area where progress has been significantly more limited is the study of nearly integrable spin chains. The specific case where integrability is broken by noisy couplings seems both numerically and theoretically tractable: quasiparticles scatter and decay at a rate set by the Golden Rule, and the main theoretical challenge is to compute the matrix elements that enter the Golden Rule. These can be calculated, or at least estimated, in many specific cases, including some with nontrivial physics like subdiffusion. For Hamiltonian perturbations that break integrability, much less is understood. In both classical and quantum systems in the thermodynamic limit, all initial conditions eventually lead to chaos, but the rate at which chaos sets in can be much slower than the naive Golden Rule estimate: this can happen, e.g., because some parts of the perturbation can be absorbed into the unperturbed Hamiltonian through a unitary rotation~\cite{PhysRevB.105.104302}~\footnote{F. M. Surace and O. I. Motrunich, private communication.}.. The prospects for studying these phenomena in near-term simulations or experiments involving quantum spin chains are not very favorable, because of the rapid growth of entanglement and the late times required. A more promising near-term avenue might be to perform large-scale simulations on nearly integrable classical Hamiltonians and develop a quantitative understanding of the effects of integrability breaking there.

\paragraph{Acknowledgments:} The authors thank  Utkarsh  Agrawal, Sounak  Biswas, Vir Bulchandani, Jacopo De Nardis, Benjamin Doyon, Michele Fava, Aaron Friedman, Paolo Glorioso,  David Huse, Enej Ilievski, Vedika Khemani, Javier  Lopez-Piqueres, Joel Moore, Vadim Oganesyan, Sid Parameswaran, Tomaz Prosen, Marcos  Rigol, Subir Sachdev,  Brayden  Ware and Marko Znidaric for collaborations and/or discussions on topics related to anomalous transport and GHD. R.V. acknowledges support from the Air Force Office of Scientific Research under Grant No.~FA9550-21-1-0123, and the Alfred P. Sloan Foundation through a Sloan Research Fellowship. S.G. acknowledges support from the US National Science Foundation under Award No. DMR-1653271. We are grateful to the KITP, which is supported by the National Science Foundation under Grant No. NSF PHY-1748958, and the program ‘Quantum Many-Body Dynamics and Noisy Intermediate-Scale Quantum Systems’, where parts of this review were finalized.

\bibliography{refs}

\end{document}